\documentclass[a4paper,fleqn]{cas-dc}

\usepackage[authoryear]{natbib}
\usepackage{longtable}
\usepackage{booktabs}
\usepackage[normalem]{ulem}
\useunder{\uline}{\ul}{}
\usepackage{enumitem}

\usepackage[most]{tcolorbox}
\newcommand{\sectopic}[1]{\vspace{0em}\par\noindent{\textit{\bfseries #1}}}

\usepackage{etoolbox}

\def\tsc#1{\csdef{#1}{\textsc{\lowercase{#1}}\xspace}}
\tsc{WGM}
\tsc{QE}
\tsc{EP}
\tsc{PMS}
\tsc{BEC}
\tsc{DE}

\begin{document}
\let\WriteBookmarks\relax
\def\floatpagepagefraction{1}
\def\textpagefraction{.001}

\shorttitle{}

\shortauthors{Vedant Chauhan et~al.}

\title [mode = title]{How do software practitioners perceive human-centric defects?}                      



%
\author[1]{Vedant Chauhan}[type=,
                        auid=,bioid=,
                        prefix=,
                        role=,
                        orcid=]



\ead{vedant.chauhan@monash.edu}



\affiliation[1]{organization={Faculty of Information Technology, Monash University},
    addressline={Wellington Road}, 
    city={Clayton},
    postcode={3800}, 
    state={Victoria},
    country={Australia}}

\author[1]{Chetan Arora}
\ead{chetan.arora@monash.edu}

\author[2]{Hourieh Khalajzadeh}[%
   role=,
   suffix=,
   ]
\ead{hkhalajzadeh@deakin.edu.au}


\affiliation[2]{organization={School of Information Technology, Deakin University},
    addressline={221 Burwood Highway}, 
    city={Burwood},
    postcode={3125}, 
    state={Victoria},
    country={Australia}}

\author%
[1]
{John Grundy}
\ead{john.grundy@monash.edu}





\begin{abstract}
Context: Human-centric software design and development focuses on how users want to carry out their tasks, rather than making users accommodate their software. Software users can have different gender, age, culture, language, disabilities, socioeconomic status, and educational background, among many other differences. Due to the inherently varied nature of these differences and their impact on software usage, preferences and issues users face can vary, resulting in user-specific defects, that we term as \emph{`human-centric defects' (HCDs)}. 

\noindent Objective: This research aims to understand the perception and current management practices of such human-centric defects by software practitioners, identify key challenges in reporting, understanding and fixing them, and provide recommendations to improve HCDs management in software engineering.

\noindent Method: We conducted a survey and interviews with software engineering practitioners to gauge their knowledge and experience on HCDs and the defect tracking process.

\noindent Results: We analysed fifty (50) survey and ten (10) interview responses from SE practitioners and identified that there are multiple gaps in the current management of HCDs in software engineering practice. There is a lack of awareness regarding human-centric aspects causing them to be lost or under-appreciated during software development. Our results revealed that the handling of HCDs could be improved by following a better feedback process with end-users, a more descriptive taxonomy, and suitable automation.

\noindent Conclusion: HCDs present a major challenge to software practitioners, given their diverse end-user base. In the software engineering domain, research on HCDs has been limited and requires effort from the research and practice communities to create better awareness and support regarding human-centric aspects.


\end{abstract}


\begin{highlights}
\item We describe the phenomenon of ``human-centric defects'' in software that are caused by end user human differences
\item We describe results from a survey of 50 practitioners and follow-up detailed interviews of 10 practitioners on human-centric defect definition, occurrence, reporting and fixing issues in software engineering
\item We identify a range of recommendations for researchers and practitioners to improve the practice of human-centric defect reporting and fixing
\end{highlights}


\maketitle


\section{Introduction}
\label{sec:intro}

Software bugs, or \emph{defects},~\citep{strate2013literature,yusop2016reporting} are a well-known phenomenon in software engineering (SE). Such defects can appear as a result of mistakes by developers, mis-understood requirements, technical problems with designs and code, or the varied nature of software end users themselves i.e., some defects are in the eye of the beholder. 

Human-centric design and development focus on creating software by placing end-user requirements at the forefront of developers' focus and effort~\citep{winograd1997challenge}. The design and development activities are centred around how end-users perceive a particular task rather than making the users adapt their behaviours to the developed software. Ideally, software requirements need to consider many differences in end users, for example, gender, age groups, cultures, languages, disabilities, socioeconomic statuses, and educational backgrounds~\citep{grundy2020human,grundy2020towards,fazzini2022characterizing}. The subjective nature of these differences means that problems can vary for individual users~\citep{khalajzadeh2022supporting}, resulting in defects in the application that we term \emph{`human-centric defects' (HCDs)}. 
Consequently, it is essential to recognise, appreciate and handle such HCDs responsibly by developers, teams, and organisations for the overall success of a product. 

What we term as conventional ``technical defects'' -- defects in software not dependent on end user human differences -- typically arise from missing functionality, developer mistakes, or unclear business requirements. Defects are often classified into (i)~functional and (ii)~non-functional defects. Functional defects violate the working of the system (i.e., functional requirements) while non-functional defects occur when the quality of the system is violated (i.e., non-functional requirements)~\citep{matrinnfr}, e.g., usability, security, performance, and compatibility defects. HCDs may fall under both these defect categories. For instance, while booking an Uber\footnote{https://www.uber.com}, the option to include the child seat is very confusing and requires extra effort from the user side to get it included in their ride. It is further not available in all locations and the option to include a car seat varies within a city. 
For a user travelling without a child under a certain age, this is not an issue. However, this is an HCD for a parent. This can be classified as a functional defect for the application e.g., no option to book a child seat. An example of a non-functional defect is when colour blindness of the parent impacts the accessibility of booking this feature.


Software defects are usually reported to developers via a \emph{defect report}. Reporters, i.e., testers or end-users, provide these reports to developers. These reports generally have several feedback fields, for example, description, expected and actual behaviour, impact, severity, and supporting supplements~\citep{bettenburg2008makes,zimmermann2010makes}. If the report feedback is presented to the developers in a tailored way, it can help solve the defects faster~\citep{strate2013literature,yusop2016reporting}. However,  research shows that defect reports are frequently incomplete and ambiguous, and the quality varies widely due to inexperience in defect reporting, communication issues, and incomplete defect form fields~\citep{bettenburg2008makes,yusop2016reporting,yusop2017analysis}. The quality of the information in defect reports can be affected due to a lack of knowledge of the bug reporting process, tools and their experience in reporting issues~\citep{ko2010power}. The information provided in defect reports often lacks clarity, or detailed information \citep{davies2014inbug}. Additionally, the information presented in many defect reports has a shortage of recommendations about defect solutions~\citep{hornbaek2006kinds} and an absence of logical and consistent textual information and stack traces~\citep{strate2013literature}. The mismatch in terms of what is required by the developers and what is presented to them further impacts the report quality~\citep{bettenburg2008makes,yusop2016reporting}.

\begin{figure}
	\centering
		\includegraphics[scale=.35]{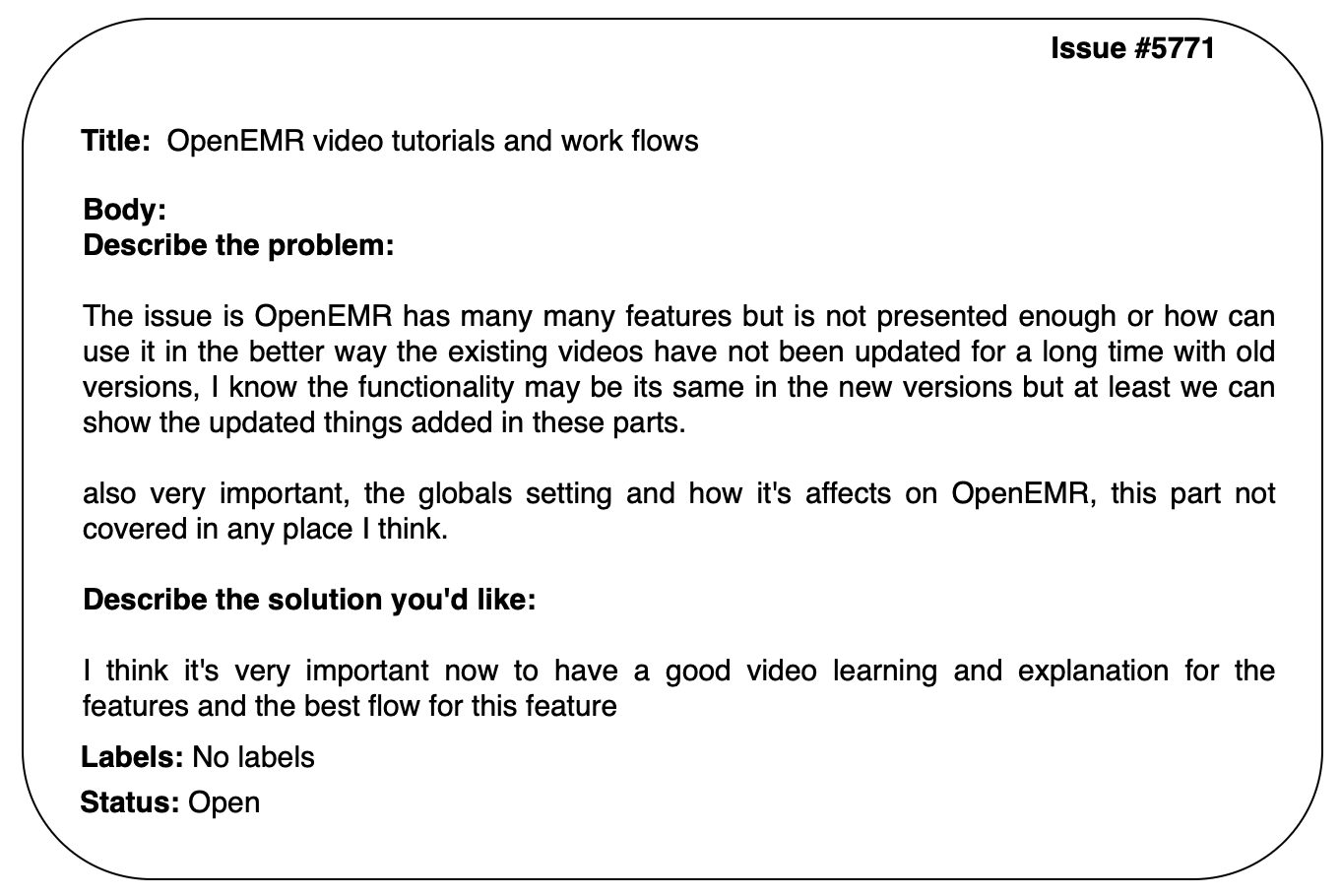}
	\caption{An example HCD reported in the OpenEMR GitHub Repository}
	\label{fig:hcdeggithub}
\end{figure}

HCDs are very nuanced and are often very subjective defects that occur when different end-users interact with the application. These defects occur due to end-user perception or differences, such as their educational background, age, gender, physical or mental disability, language, experience, and ethnic background. These defects can be considered outliers or edge conditions missed by the development teams while developing or testing their products. Figure~\ref{fig:hcdeggithub} illustrates a real-world example of an HCD reported in a healthcare open-source project called `OpenEMR'\footnote{\label{note1}https://github.com/openemr/openemr/issues/5771}. This issue highlights the problem faced by an end-user where the existing videos or materials of the application are outdated. Hence, causing user problems in utilising the features of the application in the best possible sense. For that purpose, the reporter is asking to improve the videos and explanations in the application. Overall, this could also be an issue for a wider audience of the application. The issue report contains only the basic description of the problem and a solution required by the user. There are missing defect report fields such as expected behaviour, screenshots, and additional context regarding the issue. Triaging the issue can be difficult due to specific missing information regarding videos or materials. Additionally, this issue could be a false positive for more experienced users of the application. Hence, it becomes an HCD in the usability defect category. 



Existing SE approaches either do not consider human-centric aspects or address them in an ad-hoc manner \citep{huynh2021improving,khalajzadeh2022supporting,prikladnicki2013cooperative,curumsing2019emotion}. As a result, many critical human-centric aspects are generally not collected, planned for, or addressed while creating or evaluating software ~\citep{huynh2021improving,yusop2016reporting}. Furthermore, some reporters do not effectively use defect reporting tools to capture defect information because of the lack of functionality provided by the tools to capture human-centric defect information~\citep{yusop2016influences,huynh2021improving}. Hence, fixing human-centric defects is impacted due to the quality of the resultant defect reports. There is a long-standing need for a defect taxonomy due to a lack of distinguishability and incompleteness~\citep{keenan1999usability}. 
However, human-centric aspects still require responsiveness and specificity to be handled by development teams. Teams should be able to handle HCDs responsively and tend to the specific human aspects related to the defect~\citep{huynh2021improving}.
Our research focuses on the subjectivity of defects formed due to user perception or underlying conditions, and extends to HCDs across functional and non-functional defect categories. 
To the best of our knowledge, this is the first research dealing with human aspects across functional and non-functional defect categories as well as the perception of software practitioners regarding such HCDs.

Due to the challenging characteristics of the human-centric defect reporting process and a lack of awareness about human-centric defects, the goal of this work is to understand the perception of human-centric defects in the software industry. To achieve this, we conducted a survey with 50 respondents and interviews with 10 participants. Based on our analysis of these responses, the key contributions of this paper are:
\begin{itemize}
    \item an overview of how HCDs are currently perceived by software developers and reporters during the development, testing, and maintenance phase of the SDLC;
    \item  exploration of current defect reporting tools, their challenges and enhancements required to better support HCDs; and
    \item  identification of improvements needed in managing HCDs, i.e., reporting and fixing HCDs.
\end{itemize}

The rest of the paper is structured as follows. Section~\ref{sec:relatedwork} positions our work against the related work. Section~\ref{sec:rq} discusses the research questions of this study. Section~\ref{sec:rm} provides an insight into our research methodology, i.e., our survey design and the interview process. Results and discussion from our survey and interviews are provided in Sections~\ref{sec:results} and \ref{sec:discussion}. Section~\ref{sec:threats} presents threats to the validity, and Section~\ref{sec:conclusion} concludes the paper.


\section{Related Work}
\label{sec:relatedwork}

Numerous studies have been conducted on software defects, defect triaging, remediation, reporting, quality of defect reports, automatic detection and fixing. \citet{strate2013literature} conducted a systematic review of software defect reporting targeting five potential spaces of defect tracking: triaging, quality, metrics of defect reports, automatic defect fixing and defect detection. Triaging is usually a time-consuming activity, and the SLR identifies automatic assignments, detects duplicate reports, and the validity of reports. The quality and metrics of defect reports are also a driving factor in the swiftness of fixing the issue, and SLR manages to identify the gaps in the area. Automatic defect detection and fixing focus on the manual effort in the process and what advancements can be conducted in the domain for faster fixing of defects. 

\citet{bettenburg2008makes} and \citet{zimmermann2010makes} received responses from developers and reporters revealing that bug reports suffer from poor quality because of missing or incorrect information. According to developers, steps to reproduce, stack traces and test cases are the most helpful information in defect reports, which are usually complex for the users to provide. The research developed a prototype called CUEZILLA to measure and improve the quality of bug reports by recommending elements to include in reports. Consequently, the study suggests improving four areas for bug reports: tool-centric, information-centric, process-centric, and user-centric by engaging the user and enhancing the tool support. If the quality of reporting structure is improved, it can lead to better defect management. Software defect reporting space is a widely explored domain. However, HCDs are still a domain that requires detailed investigation in relation to the whole defect-tracking process.

Several studies have been conducted to capture age, gender, personality, behaviour, culture, and physical \& cognitive disabilities. 
\citet{cruz2015forty} conducted a systematic mapping study to gauge the effects of personality on team performance. The study found considerable differences in team performance due to personality and suggested the field is not mature and requires more research. The SLR presented by \citet{soomro2016effect} explored developers' personalities and team climate impact on software performance. The study found limited research in the area and identified no agreed team climate composition in the domain of SE. \citet{burnett2016gendermag} and \citet{mendez2019gendermag} introduced techniques to evaluate gender and diversity in software development practices. They developed a gender and inclusive magnifier to generate gender and diversity insights into the product evaluation and increase the usability and overall success of the product. Developers' emotions also play a significant role in the success of a product's growth; positive emotions can improve a developer's productivity, and negative emotions can lead to frustration \citep{muller2015stuck}. The study observed that training and code review could enhance the results. \citet{lee2021persona} analysed accessibility issues under the context of home appliance usage for the target user groups: visually, hearing, spinal cord impaired, and elderly users. 

Human aspects and human-aspect-related defect studies \citep{grundy2018vision,grundy2020human,grundy2020humanise,huynh2021improving,hidellaarachchi2021effects,ahmad2023requirements} indicate the challenges in getting access to diverse users, capture and description of human-centric issues, absence of adequate feedback to developers, lack of adoption strategies, effort, and sustainability. 
More research is required to create better awareness regarding human aspects in SE \citep{grundy2020humanise}.

Several relevant studies have been carried out in the usability and accessibility domains. For instance,  some studies \citep{Bigham2010Accessibility,mbipom2011interplay,puzis2012web,feiner2016convenient,moreno2020easier} focus on accessibility problems in the website design elements for individuals with low vision, colour-blindness, visually impaired, and cognitive issues. 
Web developers who create web applications with accessible features usually fail to include usability for disabled individuals. \citet{Bigham2010Accessibility} provided an Accessibility By Demonstration (ABD) solution, which lets visually-impaired users record accessibility problems.  \citet{mbipom2011interplay} addressed the design elements in the web context and their interplay with accessibility. The study analysed what types of web pages are aesthetically pleasing or displeasing and examined the barriers to accessibility, such as low vision, colour-blindness, and visual impairment. \citet{puzis2012web} presented the automation tools used in recording and replaying web interfaces based on the history of browsing actions. 
The system enhances the understandability and readability of textual content for cognitive, language and learning disabilities. 

\cite{yusop2016reporting,yusop2016influences,yusop2017analysis,yusop2020revised} explored usability defects, the defect reporting process and the tools used for such purposes. Usability defects arise when a user experiences inadvertent behaviour while using an application. These defects are reported by expert or non-expert users. The research evaluates reporting \& challenges of usability defects and analyses usability defect information in reports. 
 Their work  identified the reporting mechanisms used for reporting and tracking issues, such as tool-based, model-based, and end-user-based. Additionally, the format and common attributes such as description, severity, steps to reproduce, stack traces, and expected behaviour, are helpful to solve defects faster. Recommendations to improve the reporting and fixing of usability defects include new taxonomy, guidelines, training for non-expert users, automation in reporting and fixing, and capturing the right attributes to provide the most applicable information regarding the defects. However, there needs to be more relevant research such as these to capture all forms of human-centric software defects. Hence, our emphasis is on improving the current reporting and fixing of HCDs in the SE domain.



\section{Research Questions}~\label{sec:rq}
Our main objective in this study is to understand the occurrence and management of HCDs in current software defect reporting and fixing processes. 
We formulated the following key research questions (RQs):\\

\sectopic{RQ1. What is the state of the practice for managing human-centric defects in software engineering?}
We wanted to investigate the current practices around reporting and handling of HCDs by software teams. We aim to understand any differences between the handling of technical defects versus HCDs. \\

\sectopic{RQ2. To what extent do software practitioners' human aspects -- such as experience, culture, emotions, language, ethnicity, gender, and age -- affect reporting and fixing HCDs?}
RQ2 gauges the significance of an individual's  traits in identifying and fixing HCDs. For instance, we posit that a practitioner's experience can impact the effectiveness of identifying and fixing HCDs. 

\sectopic{RQ3. What is the role of organisations and teams in addressing HCDs?}
 An essential factor in understanding the HCDs management process is the organisation and team's impact on identifying and fixing HCDs. For instance, how much influence do organisations and teams play in representing a human-centric approach, and do they create awareness and necessary discussions regarding human-centric aspects?\\ 





\sectopic{RQ4. What role do defect tracking tools play in reporting and fixing HCDs?}

\textit{RQ4a. What are the key challenges for defect tracking tools for reporting and resolving HCDs?}
Do defect reporting tools have any challenges in capturing information about reported HCDs.

\textit{RQ4b.What characteristics make defect tracking tools user-friendly?}
What are practitioner recommendations for improving such defect tracking tools.

\textit{RQ4c. What are the most prominent defect tracking fields required in a defect reporting form to manage HCDs better?}
To improve the HCDs reporting, we want to understand what fields are required to describe a HCD so that it can be fixed by a developer quickly.

\textit{RQ4d. Does experience with defect tracking tools influence the quality of HCD reports?}


\sectopic{RQ5. What improvements can be applied to effectively assist in reporting and resolving of HCDs?}
Finally, RQ5 investigates the improvements in resolving and reporting HCDs. This RQ provides suggestions for better management of HCDs in the SE domain.


\section{Research Methodology}
\label{sec:rm}

We conducted a survey and followed this with detailed interviews with SE
practitioners to understand their perspectives on defining, reporting and fixing HCDs. 
Sections~\ref{sec:survey} and~\ref{sec:interview} explain the design process of our survey and interviews. Section~\ref{sec:dataanalysis} defines  our data analysis process.

\subsection{Survey Process: Design and Questionnaire}
\label{sec:survey}

\begin{figure*}
  \centering
  \includegraphics[width=0.7\textwidth]{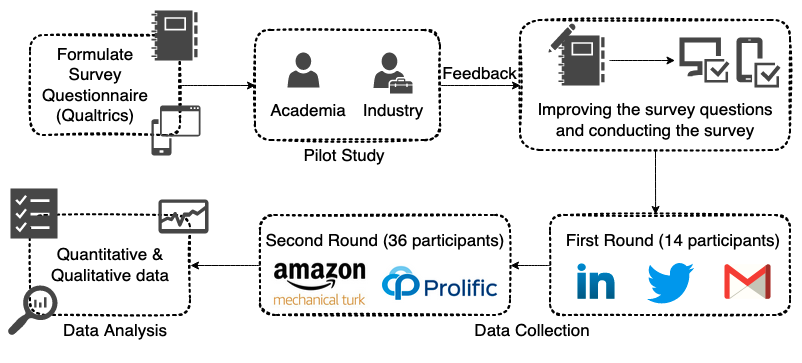}
\caption{Survey process}
\label{fig:surveyProcess}       
\end{figure*}

Figure~\ref{fig:surveyProcess} illustrates our survey design process using the Qualtrics\footnote{https://www.qualtrics.com/} platform. We constructed our questions based on the Kitchenham et al. principles \citep{kitchenham2002preliminary,kitchenham2002principles}, authors' experience in the IT industry as well as from multiple studies on defect reporting, industry practitioners' and human-centric aspects \citep{bettenburg2008makes, yusop2016reporting, yusop2016influences,votipka2018hackers,yusop2020revised, huynh2021improving, hidellaarachchi2021effects}. The total number of questions in the survey was 51, divided into five sections, and the time to complete the study was around 15-20 mins on average. We grouped our survey into five sections: Demographics, Selection, Developer, Reporter, and Conclusion:
\begin{itemize}
    \item \textbf{Demographics.} Collects  basic demographic information of the participants. The survey was anonymous, and we did not record any identifying information of participants.
    \item \textbf{Selection.} Provided participants with a choice based on their previous or current experiences. If the participants had experience developing and working on fixing the defects, they could select the developer section, and the participants were presented with only developer-related questions. Similarly, if the participants had experience in reporting defects, then they could select the reporter section, and the participants were presented with only reporter-related questions. If the participants had experience in both, they could select both. 
    \item \textbf{Developer and Reporter.} The third and fourth sections ask about how developers and reporters perceive HCDs. A first part asked about perceptions of HCDs, and a second part regarding HCD management and defect tracking tools.
    \item \textbf{Conclusion.} Asked for feedback regarding the survey and inclination towards a follow-on interview. 
\end{itemize}

Our survey questionnaires are \textbf{available on Zenodo\footnote{https://zenodo.org/}, please see~\cite{anonymous_survey_and_interview}} for more details.

After designing the questionnaire, we conducted a pilot study with industry and academic participants to validate our survey clarity, suggestions on the improvement of the survey, and time to complete the survey~\citep{baltes2022sampling}. We presented the study to four industry practitioners and one academic with prior experience in software development and reporting. We added some examples and definitions to the questions based on the feedback to provide more transparency. 

We ran our survey in two rounds. Each round was conducted over two months. First-round was conducted from Dec 2021 to Jan 2022 and was advertised on social media platforms such as LinkedIn\footnote{https://www.linkedin.com/} and Twitter\footnote{https://twitter.com/}, through the authors' colleagues, industry contacts, and via email. We received 40 responses, of which we selected 14 valid responses for further analysis after a preliminary check. Unfortunately, not-selected responses consisted of empty or incomplete responses for closed-ended questions, responses with a standard repeated answer for multiple questions, or the users who did not provide consent. 
In the second round, we decided to use crowdsourcing platforms for survey collection, such as AMT\footnote{https://www.mturk.com/} and Prolific\footnote{https://www.prolific.co/}, where we paid participants. We paid 6.40 USD/hour on AMT and, on average, 7.12 GBP/hour on Prolific for each response. This round was conducted from Feb 2022 to Mar 2022. We received 64 responses, out of which we selected 36 valid responses for data analysis. Similar to the first round, the not-selected responses consisted of empty or incomplete responses for closed-ended questions, responses with a standard repeated answer for multiple questions, or the users who did not provide consent.

\subsection{Interview Process: Design and Questionnaire}
\label{sec:interview}

\begin{figure*}
\centering
  \includegraphics[width=0.7\textwidth]{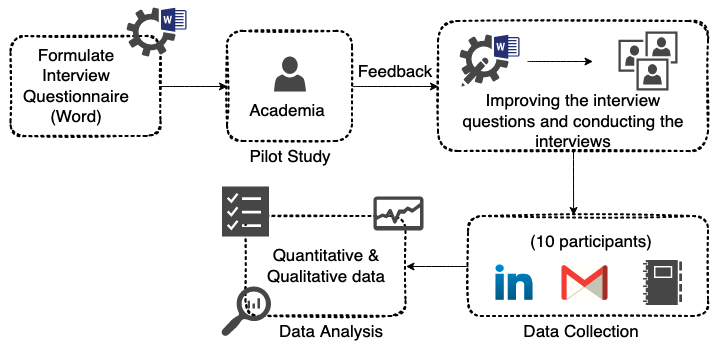}
\caption{Interview process}
\label{fig:interviewProcess}       
\end{figure*}

Figure~\ref{fig:interviewProcess} illustrates our interview design process. We constructed our questions based on the Kitchenham et al. principles \citep{kitchenham2002preliminary,kitchenham2002principles}, author's experience in the IT industry as well as from multiple studies on defect reporting, industry practitioners' and human-centric aspects \citep{bettenburg2008makes, yusop2016reporting, yusop2016influences,votipka2018hackers,yusop2020revised, huynh2021improving, hidellaarachchi2021effects}. The total number of questions in the interview was 18, divided into two sections, and the time to complete the interview was around 30 mins. We grouped our interview study in two areas: demographics and discussion of HCDs. An overview of the interview questions is as follows:
\begin{itemize}
    \item \textbf{Demographics.} The first section contained participants' demographic information. The interview was anonymous, and we did not record any personally identifiable information of participants. After recording the session, we transcribed the interview and deleted the recordings.
    \item \textbf{HCDs Discussion.} The second section contained a semi-structured set of questions designed to keep the conversation going when discussing the themes such as HCDs examples, the influence of emotions and culture, defect reporting, improvements, taxonomy, and automation related to human aspects.
\end{itemize}

Our interview questions are \textbf{available on Zenodo, please see ~\cite{anonymous_survey_and_interview}} for more details.

After designing the interview questionnaire, we again piloted the study with industry-experienced academic researchers to validate our interview questions. 
We then conducted our interview study after the second round of the survey. We had two motivations to conduct the interview study. First, the survey responses were limited and second, surveys are often restrictive in providing thorough responses to some key questions~\citep{couper2000web}. In our case, we wanted to understand more about current industry practices impacting HCDs. We did not have dedicated sections on the developer and reporter profiles, but the second segment of our interviews focused on covering HCDs both from the developers' and reporters' perspectives.


We conducted the interviews from Apr 2022 to May 2022. 
Of the ten professionals interviewed, only four had not previously completed the survey and were industry contacts of the authors. Post interviews, we provided our interview participants with a choice of a 20\$ gift voucher in their currency from Myers\footnote{https://www.myer.com.au/}, Coles\footnote{https://www.coles.com.au/}, iTunes\footnote{https://www.apple.com/au/itunes/}, and Amazon\footnote{https://www.amazon.com/} for their time and participation in the interview study.

\begin{figure}
\centering
  \includegraphics[scale=.3]{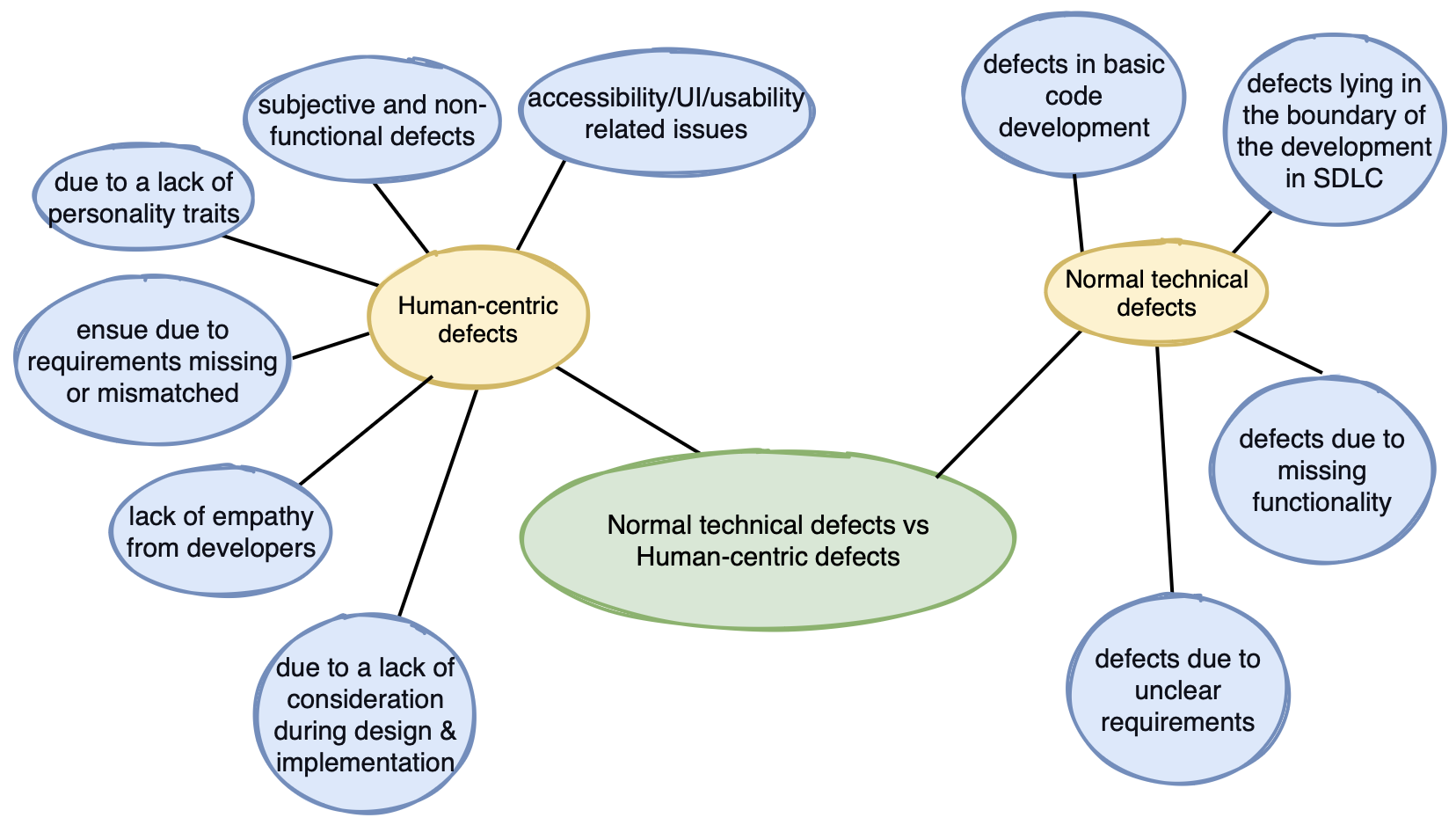}
\caption{Theme map for RQ1}
\label{fig:thememap}       
\end{figure}

\subsection{Data Analysis}
\label{sec:dataanalysis}

We collected both qualitative and quantitative data using surveys and interviews. First, we analysed survey responses. Based on the analysis we identified general themes which were spread across interviews. Since the interviews were a deeper extension of the survey answers tackling more prominent and detailed themes, we decided to combine the analysis of both and generate combined themes from them. 
We utilised descriptive statistical analysis on the quantitative data using the Qualtrics platform, as shown in Section~\ref{sec:demographics}. 

For the qualitative data, we used thematic synthesis \citep{cruzes2011recommended}.  The thematic analysis identifies, analyses, and reports patterns or themes within qualitative data \citep{braun2006using}. However, if the thematic analysis is not applied inside an underlying conceptual framework, it has minimal explanatory ability beyond basic description \citep{braun2006using}. Hence, we decided to use thematic synthesis. Thematic synthesis builds on the principles of thematic analysis and identifies, explains, and summarises the themes captured in the process. We utilised this concept to determine the recurring patterns in the qualitative data and establish the purpose and summary of those patterns for our overall research questions. After analysing the qualitative data, we identified specific text segments in the data. We then labelled the text using an open coding technique to build ideas, concepts, and overall themes~\citep{khandkar2009open}. Figure~\ref{fig:thememap} provides a sample theme map for RQ1 (see~Section~\ref{sec:rq1}). It shows all codes for the theme of HCD and normal technical defects. Similarly, we used the open-coding technique for open-ended questions by constant comparison, creating lower-order themes and merging them into higher-order ones.


\section{Results}
\label{sec:results}


\subsection{Demographics}
\label{sec:demographics}

Survey respondents are referred to as \emph{SR-x} and interview participants \emph{IP-x}. When referring to both respondents and participants in this paper, we refer to them generally as practitioners.

\subsubsection{Survey}
\label{sec:demographicssurvey}

\begin{table*}[]
\caption{Survey participants' demographics information}~\label{table:surveydemographics}
\footnotesize
\begin{tabular}{@{}ll@{}}
\hline\noalign{\smallskip}
\textbf{Demographic Information of the participants}                                                               & \textbf{Overall} \\ \hline\noalign{\smallskip}
\textbf{Countries of the participants}                                                                             & \textbf{}        \\ \hline\noalign{\smallskip}
Australia, the UK, Mexico, Poland, and South Africa                                                                & 50\%             \\
Chile, Portugal, USA, and Italy                                                                                    & 24\%             \\
Spain, Romania, Canada, Netherlands, \\
Iran, Greece, New Zealand, Austria, and Denmark                               & 18\%             \\
Sweden and India                                                                                                   & 8\%              \\ \hline\noalign{\smallskip}
\textbf{Age ranges of the participants}                                                                            & \textbf{}        \\ \hline\noalign{\smallskip}
25-29                                                                                                              & 32\%             \\
30-34                                                                                                              & 20\%             \\
40-50                                                                                                              & 18\%             \\
20-24                                                                                                              & 14\%             \\
35-39                                                                                                              & 6\%              \\
51-60                                                                                                              & 6\%              \\
60+                                                                                                                & 4\%              \\ \hline\noalign{\smallskip}
\textbf{Gender of the participants}                                                                                & \textbf{}        \\ \hline\noalign{\smallskip}
Male                                                                                                               & 60\%             \\
Female                                                                                                             & 40\%             \\ \hline\noalign{\smallskip}
\textbf{Highest Education}                                                                                         & \textbf{}        \\ \hline\noalign{\smallskip}
Bachelor's degree                                                                                                  & 44\%             \\
Master's degree                                                                                                    & 32\%             \\
High school graduate, diploma, or \\
the equivalent and Some college credit, no degree                                 & 12\%             \\
Doctorate degree                                                                                                   & 4\%              \\
Trade/technical/vocational training                                                                                & 4\%              \\
Associate or Professional degree                                                                                   & 4\%              \\ \hline\noalign{\smallskip}
\textbf{Education information of the participants}                                                                 & \textbf{}        \\ \hline\noalign{\smallskip}
Computer Science, Information technology, \\
Computer Engineering, Data Science, 
Programming, and Information Systems & 86\%             \\
Arts and Humanities, Business, Education, \\
Engineering Physics, and Civil Engineering                               & 14\%             \\ \hline\noalign{\smallskip}
\textbf{Job roles of the participants}                                                                             & \textbf{}        \\ \hline\noalign{\smallskip}
Software Engineer/Software Developer/Programmer                                                                    & 48\%             \\
Operations, Consultant, Database Engineer, \\
UI/UX, and Security Engineer                                            & 20\%             \\
Project Management, Business Analyst, and Partner Manager                                                          & 14\%             \\
Test Engineer/Tester/Quality Assurance                                                                             & 10\%             \\
Delivery Engineer/Lead                                                                                             & 8\%              \\ \hline\noalign{\smallskip}
\textbf{Total Experience of participants in IT industry}                                                           & \textbf{}        \\ \hline\noalign{\smallskip}
3-5 years                                                                                                          & 34\%             \\
9+ years                                                                                                           & 28\%             \\
0-2 years                                                                                                          & 22\%             \\
6-8 years                                                                                                          & 16\%            \\ \hline\noalign{\smallskip}
\textbf{Total Experience in defect reporting}                                                           & \textbf{}        \\ \hline\noalign{\smallskip}
0-2 years                                                                                                          & 64\%             \\
3-5 years                                                                                                         & 16\%             \\
9+ years                                                                                                          & 14\%             \\
6-8 years                                                                                                           & 6\%             \\
\hline\noalign{\smallskip}
\textbf{Total Experience in defect reporting tools}                                                           & \textbf{}        \\ \hline\noalign{\smallskip}
0-2 years                                                                                                          & 67\%             \\
3-5 years                                                                                                         & 20\%             \\
9+ years                                                                                                          & 13\%             \\
6-8 years                                                                                                           & 0\%             \\
\hline\noalign{\smallskip}
\textbf{Field of Work}                                                           & \textbf{}        \\ \hline\noalign{\smallskip}
Information Technology (IT)                                                                                                          & 24\%             \\
IT-Finance                                                                                                         & 16\%             \\
IT-Healthcare                                                                                                          & 16\%             \\
IT-Manufacturing                                                                                                           & 14\%             \\
IT-Spatial                                                                                                           & 10\%             \\
IT-Gaming                                                                                                          & 8\%             \\
IT-Government                                                                                                          & 6\%             \\
IT-Telecommunications                                                                                                         & 6\%             \\
\hline\noalign{\smallskip}
\end{tabular}
\end{table*}

\begin{figure}
\centering
  \includegraphics[scale=.35]{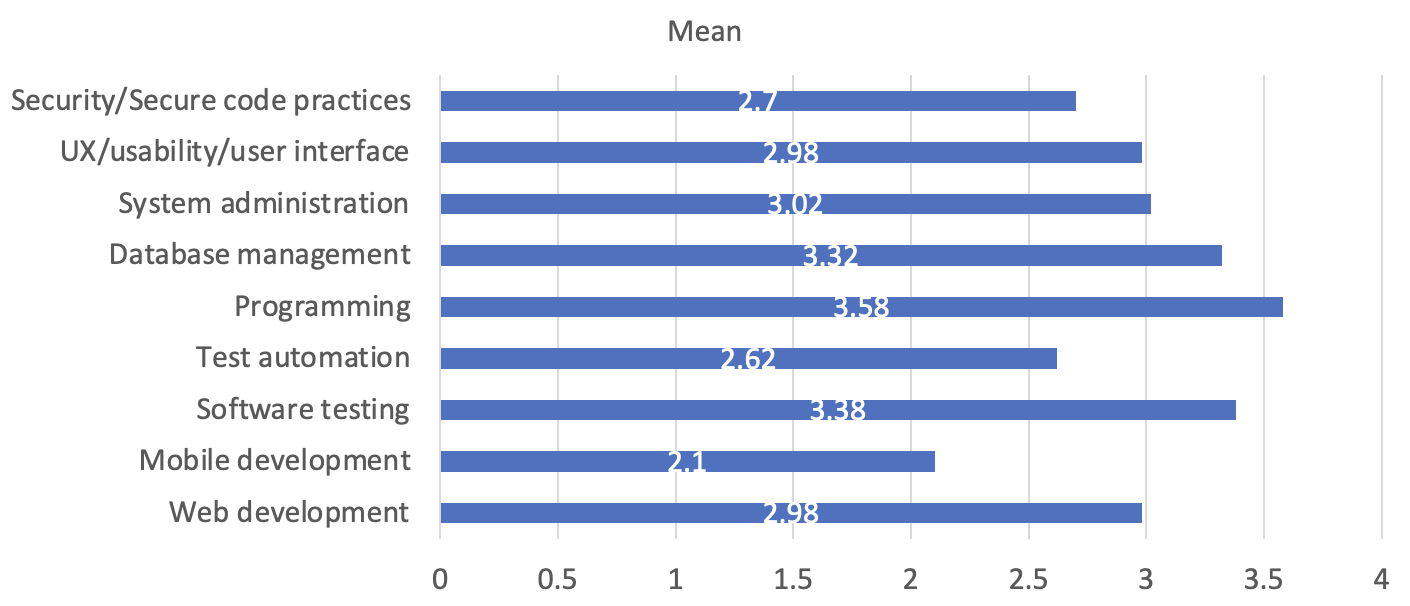}
\caption{Mean plot of technical skills of the respondents (in Surveys)}
\label{fig:meanplotskills}       
\end{figure}

Table~\ref{table:surveydemographics} shows the demographics of the 50 survey respondents, which were considered in our final analysis. 
Our target participants were developers and reporters. We thus wanted to gauge the technical skills and roles of the respondents in software development and testing activities. Figure~\ref{fig:meanplotskills} shows the skills measured on a Likert scale of 1 to 5, with 1 being a beginner or no experience and 5 being the expert. The scale ranges in the order of novice/no experience, advanced beginner, competent, proficient, and expert~\citep{hl1986mind}.  Figure~\ref{fig:meanplotskills} provides a mean plot of the development, testing, database, usability, administration, and security skills with the respondents’ experience in these skills ~\citep{florea2019skills}. 
The respondents had the most experience in programming, and mobile development was rated as the one with the least experience. The most common skills were web development, testing, and user interface design. 
We received 22 responses from only developers, 16 responses from only reporters, and 12 from respondents who were experienced, as both.

\subsubsection{Interview}
\label{sec:demographicsinterview}
Table~\ref{table:interviewdemographics} shows the demographics of the 10 interview participants. We received most responses from Australia (50\%), while the remaining participants were from Canada, the USA, Germany, and Romania. 
All ten participants had a university degree in Computer Science, Information Technology, Data Science, and Software Engineering. We interviewed participants with varied job roles: developers, data scientists, testers, consultants, and customer-centric innovation engineers. All the participants had previous or current experience in software development or reporting issues.

\begin{table*}[]
\caption{Interview participants' demographics information}~\label{table:interviewdemographics}
\begin{tabular}{@{}ll@{}}
\hline\noalign{\smallskip}
\textbf{Demographic Information of the participants}                                                   & \textbf{Overall} \\ \hline\noalign{\smallskip}
\textbf{Countries of the participants}                                                                 & \textbf{}        \\ \hline\noalign{\smallskip}
Australia                                                                                              & 50\%             \\
USA, Germany, and Romania                                                                              & 30\%             \\
Canada                                                                                                 & 20\%             \\ \hline\noalign{\smallskip}
\textbf{Age ranges of the participants}                                                                & \textbf{}        \\ \hline\noalign{\smallskip}
25-29                                                                                                  & 50\%             \\
30-34                                                                                                  & 30\%             \\
40-50                                                                                                  & 20\%             \\ \hline\noalign{\smallskip}
\textbf{Gender of the participants}                                                                    & \textbf{}        \\ \hline\noalign{\smallskip}
Male                                                                                                   & 90\%             \\
Female                                                                                                 & 10\%             \\ \hline\noalign{\smallskip}
\textbf{Highest Education}                                                                             & \textbf{}        \\ \hline\noalign{\smallskip}
Master's degree                                                                                        & 70\%             \\
Doctorate degree                                                                                       & 20\%             \\
Bachelor's degree                                                                                      & 10\%             \\ \hline\noalign{\smallskip}
\textbf{Education information of the participants}                                                     & \textbf{}        \\ \hline\noalign{\smallskip}
Computer Science, Information technology, 
Computer Engineering,  \\
Data Science, and Software Engineering & 100\%            \\ \hline\noalign{\smallskip}
\textbf{Job roles of the participants}                                                                 & \textbf{}        \\ \hline\noalign{\smallskip}
Software Engineer/Software Developer/Programmer                                                        & 40\%             \\
Data Scientist                                                                                         & 20\%             \\
Consultant and Customer-Centric Innovation Engineer                                                    & 20\%             \\
Test Engineer/Tester/Quality Assurance                                                                 & 10\%             \\
Data Engineer                                                                                          & 10\%             \\ \hline\noalign{\smallskip}
\textbf{Total Experience of participants in IT industry}                                               & \textbf{}        \\ \hline\noalign{\smallskip}
9+ years                                                                                               & 40\%             \\
3-5 years                                                                                              & 30\%             \\
6-8 years                                                                                              & 30\%            \\ \hline\noalign{\smallskip}
\textbf{Field of Work}                                                           & \textbf{}        \\ \hline\noalign{\smallskip}
IT-Finance                                                                                                         & 30\%             \\
IT-Healthcare                                                                                                          & 20\%             \\
IT-Government                                                                                                           & 20\%             \\
IT-Spatial                                                                                                           & 20\%             \\
IT-Manufacturing                                                                                                          & 10\%             \\
\hline\noalign{\smallskip}
\end{tabular}
\end{table*}

\subsection{RQ1 Results - Current state of the practice for managing HCDs}
\label{sec:rq1}

We asked about practitioner's views on the disparities between technical defects and HCDs, as well as defect categories being reported in the industry. 
To comprehend the management of HCDs in current industry practice, we identified the following key themes from the responses:


\subsubsection{Frequently reported HCD categories}
\begin{figure}
\centering
  \includegraphics[scale=.28]{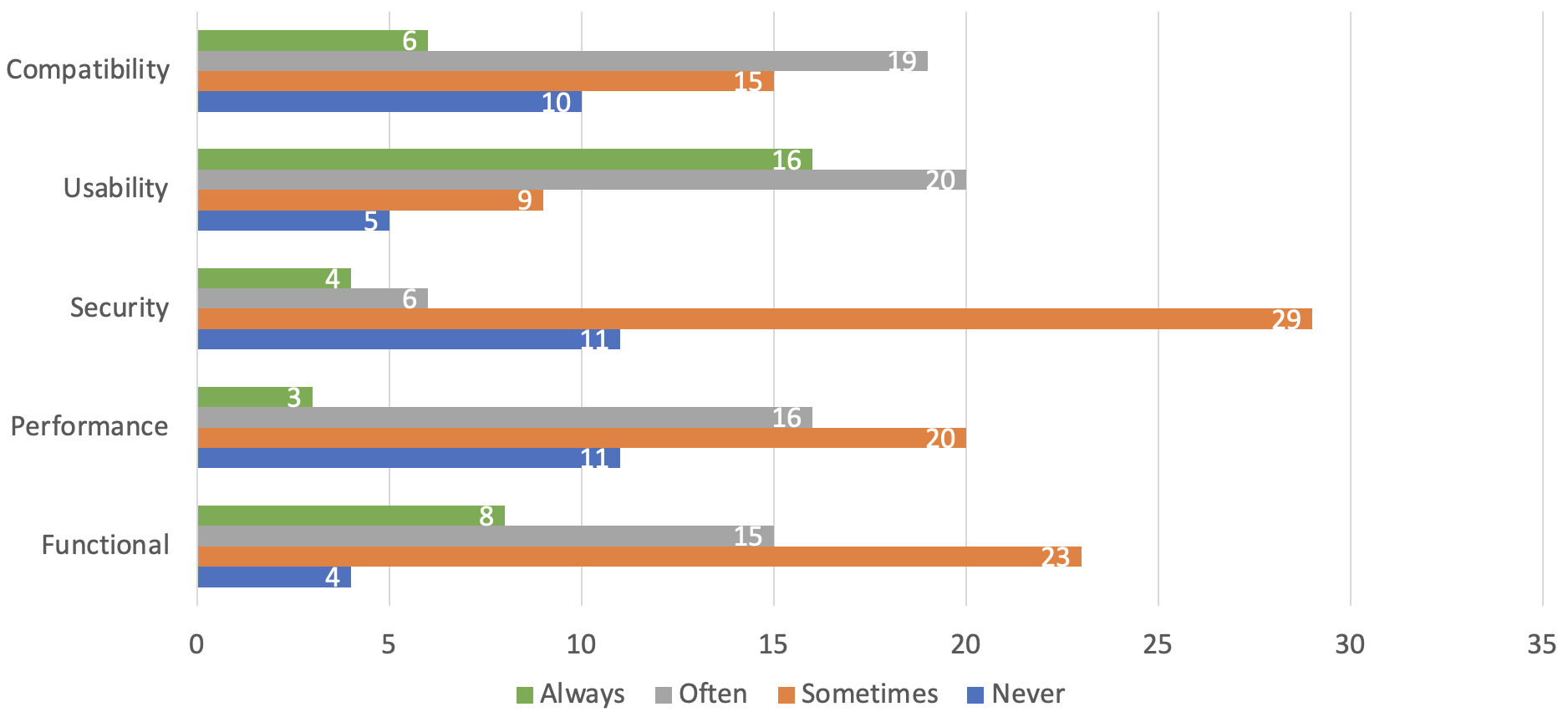}
\caption{Defect categories provided by developers and reporters (in Surveys).}
\label{fig:defectcategories}       
\end{figure}

We presented respondents with a Likert scale of \textit{`never, sometimes, often, and always'} to highlight HCDs in defect categories, such as functional, performance, security, usability, and compatibility. Figure~\ref{fig:defectcategories} shows a bar graph with the usability defect category comprising most HCDs. 
  The graph shows  
 36 responses out of 50 survey responses for the usability defect category under `always' and `often' scale. The security defect category contains the least HCDs according to the responses. Overall, the usability category is followed by functional and compatibility and concludes with the performance and security defect categories, respectively. 

Some interview participants provided examples of HCDs in functional and non-functional defect categories based on their personal experiences. For instance, an interview participant explained a usability issue in Microsoft Teams\footnote{https://www.microsoft.com/en-au/microsoft-teams/group-chat-software} and reported: \textit{``This is what I experienced daily, in my meetings (on MS Teams) - when we blur the background, it instantly recognises the face of one of my colleagues. I'm a colored person, it does not recognise my face, and it is just blank. It takes some time to recognise my face.'' - [IP-4].} 

A functional defect recognised by one of the interview participants in a software solution where daylight saving can cause developers to misunderstand when the issue occurred, \textit{``The confusion caused due to daylight saving and date formats.'' - [IP-1].}


\subsubsection{Practitioners' perception of the differences between HCDs and technical defects} From the gathered responses, normal technical defects arise for everyone during code development, due to missing functionality, or unclear business requirements. For example, IP-10 claimed: 
\textit{``Normal technical defects are the basic development defects, which come to developers when the code is not working as per the functionality or as per the business rules that have been defined''.} 
However, HCDs are subjective defects that occur due to the lack of consideration of human aspects, empathy, and missing requirements~\citep{arora2019empirical,grundy2020human,gunatilake2023empathy}. For instance, a survey respondent commented that \textit{``HCDs are biased, uncoordinated, inaccurate, or imperfect results during project development.'' - [SR-48].}

Overall, 96.7\% of our survey and interview responses state there is a similarity in technical defects and HCDs as they both arise in SDLC due to missing or unclear requirements. However, there is a difference as HCDs are perceived as more subjective and technical defects are more definitive. HCDs arise from the necessity of an individual, the philosophy of individuals, teams and organisations, and the human perspective. They are unique and hence, considered differently from normal technical defects. 
For instance, a survey respondent on HCDs mentioned: \textit{``Since each person is unique, these human-centric defects appear out of necessity, either because of the culture, type of learning, and experience of each person.'' - [SR-47].}


\subsubsection{Approaches for identification of technical defects and their application to HCDs}
\begin{figure}
  \includegraphics[scale=.4]{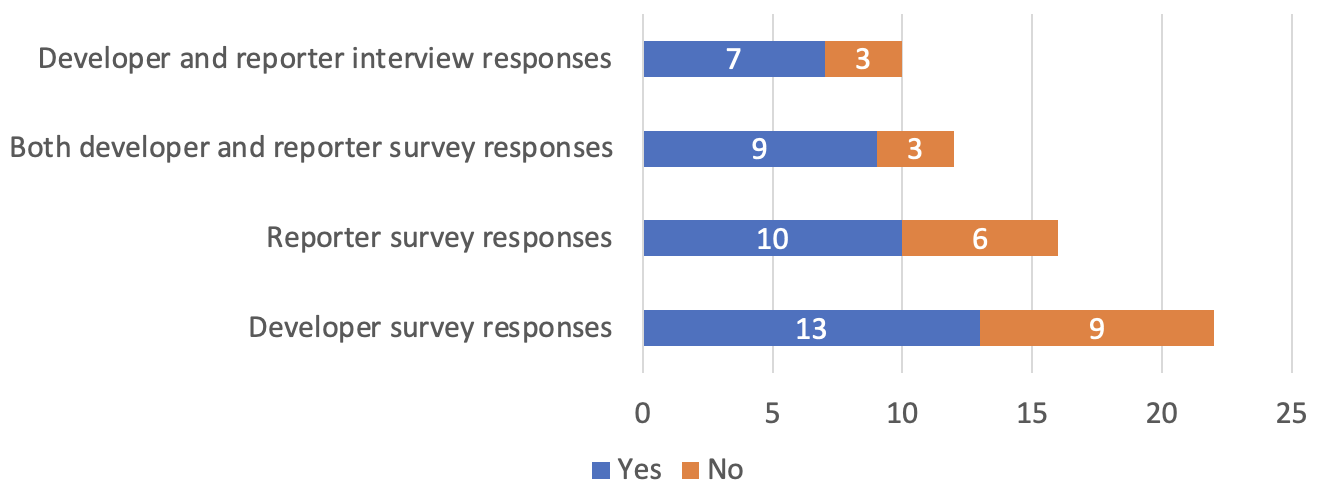}
\caption{Bar chart of HCDs experienced by developers and reporters (in Surveys and Interviews).}
\label{fig:barchartexp}       
\end{figure}

Practitioners provided their experiences in defect identification, triaging, and resolution to understand the nature of software defects and their identification. They also showcased some examples and methodologies which can be utilised for HCDs. As shown in Figure~\ref{fig:barchartexp}, over 65\% respondents experienced 
what they defined as HCDs in previous or current projects. Our respondents and participants work in different application domains, such as health, finance, manufacturing, government, telecommunications, and information technology, as shown in Tables~\ref{table:surveydemographics} and ~\ref{table:interviewdemographics}. 

Practitioners reported that they identify defects by following one or more of the following steps: (i)~reviewing bug reports,
log files,
invalid logical flows,
requirement specifications,
expected results vs. actual results, and
similar errors detected by the specialists;
(ii)~trial and error;
(iii)~categorising issues into functionality, performance and usability;
(iv)~considering how a user will perceive the problem based on needs, ease of use, accessibility and expected behaviour;
(v)~performing peer reviews;
(vi)~creating models, e.g., data models and UX flows;
(vii)~positive and negative testing of data input possibilities; and
(viii)~recreating the user path.

On defect identification, SR-3 commented that \textit{``While categorising or segregating issues, I try to divide the problems into functional, performance and usability. A lot of the human-centric issues can be addressed while writing the functional requirements. But once the application is developed, just addressing functional requirements might not resolve the problem. We might have to dive as deep as changing the fundamental architecture or database at a granular level, like entity-relationship mapping.''}

Most respondents claimed that HCDs could benefit from a similar identification process, such as developers understanding user needs and resolving issues. For example, some survey respondents shared their perspective on the identification process:
\textit{``Who, what, and why are the questions I ask while fixing an issue. Who raised the issue, what caused the issue and why the issue was present in the first place. Yes, the same questions can be applied to human-centric defects as well.'' - [SR-2].\\
``Identification is a process of trial and error, but I always try to think on how to break the design and the same could be applied to human-centric defects since you could always put yourself in someone else's shoes.'' - [SR-44].} 

\subsubsection{Software defect assignment to the practitioners and their application to HCDs}
\begin{figure}
  \includegraphics[scale=.4]{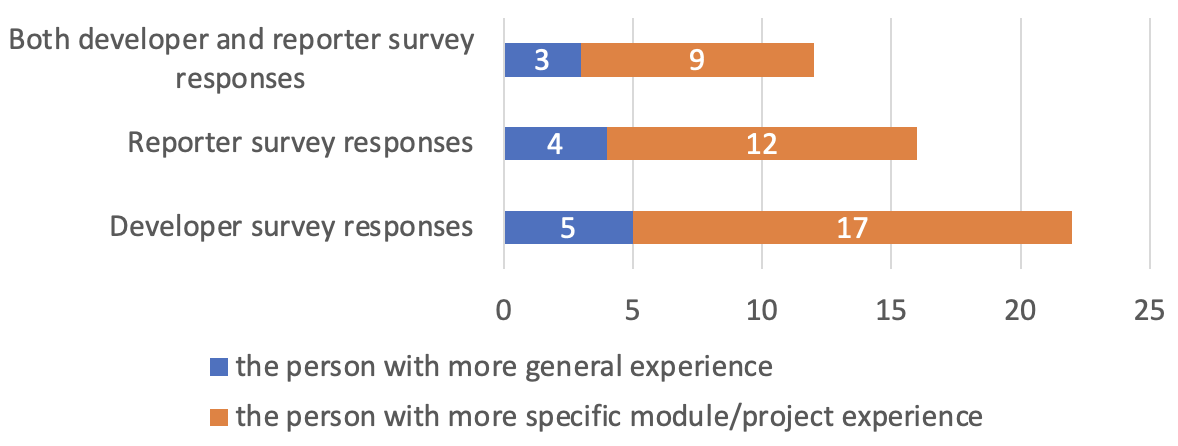}
\caption{Bar chart of defect assignment (in Surveys).}
\label{fig:barchartdefassign}       
\end{figure}

Defect assignment depends on the specific software module and general experience of practitioners. We define general experience as the total experience in the industry and the module experience as the total experience in a specific module of the project or the project itself. How particular HCDs are assigned to  practitioners is an interesting question.  
76\% of our respondents observe that a defect could be assigned to someone with more specific project experience, as shown in Figure~\ref{fig:barchartdefassign}. Respondents claimed that it is more practical to allocate defects to the individual with specific module/project expertise where the defect occurs. For example, a respondent claimed,  \textit{``Usually developers who had previously worked on a module, can resolve any errors within the module faster given their understanding of the module. Similarly, human-centric defects can be mitigated easily if developers understand the entire paradigm of those defects, their causes beforehand and formulate a solution accordingly.'' - [SR-4].} 

In our interviews, we asked for the rationale behind allocation of HCDs based on expertise. We found that organisations and teams utilise expertise to help provide the best solution in a shorter time due to the methodologies used, such as Agile. Project and resource pressure is usually the time pressure that occurs due to having to meet release deadlines and is one of the drivers in the assignment of defects to a developer. For instance, any developer who has worked on a given module will also be responsible for resolving the issues in the module within a sprint. An interview participant suggested, \textit{``Since the work is in an Agile methodology, we will first check the availability of a person. Therefore, if a defect is encountered and a person who has previous experience with the defect, but working on a different assignment should not be encumbered with the defect. However, if the defect priority is high, the person has to prioritise it above other tasks.  Generally speaking, preference will be given to the person who has already worked on the defect. For example, XYZ defect has been encountered, and there are two people A and B. Person A has previous experience with the defect and the module. Hence, Person A will be given the precedence to work on the defect.'' - [IP-10].} 

Practitioners further noted that the defects are assigned to other developers if the responsible developer is not available. For instance, an interviewee stated on defect assignment, \textit{``If someone is involved in feature development and they are not available to look at the defects, then generally whoever else is available is assigned the defect.'' - [IP-9]}. Sometimes assignment of defects also depends on the nature of the defect. If the defect falls under human aspects then the practitioner who has more experience in HCDs should be assigned the defect. For instance, some survey respondents indicated: \\
\textit{
``In general, the person who has the most experience with defect reporting process will work on the defect. For human-centric defects, it might make sense to have someone with a more general experience to take a second look.'' - [SR-30]\\
``For sure someone who has more experience on human-centric defects can be faster and better at resolving them'' - [SR-33].}

An argument was that since HCDs are often subjective, any developer can understand the defects. For instance, a survey respondent claimed, \textit{``I suppose that human-centric defects can be assigned to any developer because any developer can understand the principles of human-centric defects, while normal defects require specific module experience to be solved'' - [SR-22].}

\begin{tcolorbox}[arc=0mm,width=\columnwidth, top=1mm,left=1mm,  right=1mm, bottom=1mm, boxrule=1pt] 
\small
\textbf{RQ1 Key findings.} 
\begin{itemize}[leftmargin=*]
   \item HCDs are often reported under the usability category followed by functional, compatibility, performance, and security defects.
   \item Normal technical defects are more definitive and arise from usage, design and development. In contrast, HCDs are subjective and derive from individual differences based on different human aspects, e.g., age, culture, language proficiency, emotions, and experience.
    \item Software practitioners identify technical defects by looking at bug reports, log files, logical flows, trial and error, categorising issues, checking requirement specifications, and checking functionality, performance, and usability of the application. The same approach can be applied to identify HCDs.
    \item Software defects are assigned to developers who have more specific module expertise rather than the general overall experience. If the defect falls under human aspects, then the practitioner with more experience in HCDs should be assigned the defect. The other argument is that since HCDs are subjective, any developer can understand the defects.   
\end{itemize}
\end{tcolorbox}

\subsection{RQ2 Results - Extent to which the human aspects of software practitioners affect reporting and fixing HCDs}\label{sec:rq2}

We define human aspects as the set of differences between individuals, e.g., which culture or ethnic group they belong to, their experience, emotions, educational background, gender, age, and the languages they speak~\citep{khalajzadeh2022supporting,grundy2020human}.
Based on practitioners' feedback, we identified that human aspects are an influential factor in how developers or reporters perceive human-centric defects in general. The key human aspects identified by our survey and interviews include:
(i) knowledge and experience of an individual; 
(ii) communication and language;
(iii) emotional background (individuals experiencing emotions which reflect their mental state \citep{jamesemotions} is termed as `emotional background' of an individual);
(iv) diversity, ethnicity, culture, and age.

Different human aspects of developers and reporters need to be identified and respected by leadership in organisations and teams. Individuals and teams should be flexible enough to look beyond just their own personal and organisational values and environment to identify HCDs. For example, an interview participant claimed, \textit{``If the developers are not flexible to look beyond their cultural values, the environment they work in or adapt to the conditions, they will not be able to discover human-centric defects which are global in nature'' - [IP-1].}

94\% of respondents claim that the \textbf{experience of a practitioner plays a crucial role}. 
The experience level is directly proportional to the level of understanding of a defect and the efficiency in resolving the issues. For instance, SR-27 highlighted, \textit{``Experienced workers can quickly identify complicated issues. They have the experience to notice things which others ignore''.} Experience provides the knowledge and learning to solve a variety of problems. Experienced developers and reporters can anticipate issues and provide the best possible mitigation. For instance, some survey respondents claimed: \\
\textit{``Experience is easily the best mitigation. If you work in the same field for many years, the same problems (human-centric or other) arise and you're better equipped to solve them.'' - [SR-1]\\
``While the defect context and developers context have common points, for example, same age, same language, etc. Experiences can be useful for understanding problem statement'' - [SR-7].}


A few individuals disagreed that experience helps in identifying and fixing HCDs. According to them, if HCDs are related to inclusiveness and diversity, experience may not have much of a bearing on understanding human concerns. For example, SR-3 posited: \\
\textit{``Experience may or may not help in fixing human-centric defects. For example, if the experience has been in an environment full of people from diverse regions/countries/cultures, we can learn about human-centric issues and fix them. At the same time, experience with people from a limited regional area might not provide good insights regarding human-centric defects.'' - [SR-3].}

A vast majority of practitioners highlighted \textbf{communication as an important factor} in the understanding of HCDs from impacted end users. Developers and reporters do not directly handle client feedback in many organisations. They are often managed by business analysts or product owners and communicated down the line to developers. In some cases, there are scenarios that lead to mishandling of user requirements as people who are building software are different from those who are testing or managing. For instance, IP-1 suggested,  \textit{``A big issue is communication. In my experience, you usually have a team of people with one person as the contact point to the client, and other people who are going to be using the software. The issues need to be communicated to a range of people i.e., developers who are not on the front line, they're not sort of dealing with the client directly''.} 

Emotions are diverse phenomena that represent distinct body expressions, e.g., happy, sad, fear, disgust, and so on \citep{frijdaemotions}.  Individuals experiencing these emotions reflect their mental state \citep{jamesemotions} and hence we term it as the \textbf{\textit{`emotional background'} of an individual}. Sometimes emotions of a developer can impact the solution to an issue. Developers' mental state can be sometimes stressful while working on issues under pressure situations. For example, IP-6 claimed, \textit{``Developers usually work in pressure situations due to deadlines and constant issues with projects and working environment can lead to stress''.} Some individuals claim that knowledge and experience play a more significant role than the 
`emotional background' of a developer. For example, an interviewee claimed, \textit{``I think emotions play a very significantly less factor than the background experience. For example, experience and knowledge of your field'' - [IP-3].} As a result, this helps in providing a much better end-user experience in their software. 


Including \textbf{diversity in the development teams} can provide different perspectives and experiences to aid in understanding and appreciating HCDs. In addition, different perspectives bring new ideas and opinions that can help design, implement, and test solutions. For instance, IP-10 suggested, \textit{``Diversity in your team is always a plus point. Because it brings different points of views and thinking which when collaborated can help in achieving the best solutions to the tough questions as well''.} Exploration of different perspectives and values can be experienced when you reside in a particular community where you explore different perspectives and values which shape your knowledge on the working of the society. This can be beneficial in providing the best possible solution. For instance, IP-3 claimed, 
\textit{``...cultural diversity can help in elevating some of the issues and lead to best possible solutions''.}

\begin{tcolorbox}[arc=0mm,width=\columnwidth,top=1mm,left=1mm,  right=1mm, bottom=1mm, boxrule=1pt] 
\small
\textbf{RQ2 Key findings.} 
\begin{itemize}[leftmargin=*]
    \item Human aspects play an important role in reporting and fixing many software defects.
    \item Software practitioners’ experience plays a crucial role in identifying and fixing HCDs. 
    \item Communication and translation of issues can lead to proper handling/mishandling of user requirements, particularly those that result in HCDs. 
\item Emotions of a developer can impact the resolution of HCDs. However, the knowledge and experience of a developer are deemed more crucial in identifying HCDs than their emotions. 
    \item Diversity in the teams can provide enriching experiences and unique perspectives to design, test, and develop solutions for managing HCDs.
\end{itemize}
\end{tcolorbox}

\subsection{RQ3 Results - Role of organisations and teams in addressing HCDs}
\label{sec:rq3}

Organisations and teams want to deliver high-quality products \citep{mockus2010organizational}, thus it is essential to understand the role organisations and teams play in handling HCDs. 
Organisations and teams do not seem to handle these HCDs comprehensively. We identified the following key themes from our analysis.

\subsubsection{HCD Awareness (or lack thereof)}

First, there is a lack of awareness regarding HCDs within teams and organisations, which creates a problem when communicating these defects from project managers or business analysts to the developers. Two  interview participants reported, 
\textit{``I'm an experienced software developer having experience of eight to nine years. I have not encountered the topic of human-centred defects before two or three years ago. So it came quite late into my information that these issues also exist'' - [IP-1].} \\ and
\textit{``I think most people in organisations will not be looking into human-centric defects because first of all, the organisation will have to provide awareness regarding this'' - [IP-9].}

Secondly, due to the extensive workload and the structure of delivery and planning, there could be a possibility that organisations' and teams' prioritisation lies in satisfying the product functional deadlines and not fixing HCDs. For instance, IP-3 reported, \textit{``I think planning can solve these issues. However, sometimes on short deadlines, the time for planning is less and then human-centric problems can arise frequently. Therefore, it becomes an ongoing cycle of rectifying these problems''.} There should be some change management from the organisations to highlight HCDs; otherwise, developers will solely focus on the code and release of the product before the deadline. A customer-centric process for finding defects should be undertaken to highlight HCDs and resolve them faster. IP-6 highlighted this by saying \textit{``...there should be some change management from the organisation to highlight human-centric defects better otherwise, a developer will only focus on releasing the code. There should be a customer-centric thought process for finding these defects better''.}

Similar to our findings in RQ2, practitioners highlighted the importance of diversity in teams and organisations. They emphasised that  
\textit{``Diverse teams will come up with different scenarios and ways in which people will use the product because of their upbringing and environment'' - [IP-5].} \\
\textit{``In case the organisation is an IT consulting firm or develops multiple software products for clients, the company can set up a community of reviewers from a diverse group. Human-centric defects should be listed and shared with all so that similar issues do not occur repeatedly'' - [SR-3].}

\subsubsection{Organisation's size impacting HCDs}

A large majority of respondents (84\%) believe that an organisation's size impacts the handling of HCDs. Of these, there was a balance in terms of the  responses on the advantages and disadvantages of a small and large organisation handling HCDs. For instance, SR-1 claimed, \textit{``Plus and minus for both: bigger organisation has more resources to identify/address the problem, but requires more communication (chain of fixers are often further away from end-user experiencing the issue). I think these issues balance each other out''.}

Large organisations can be more bureaucratic, take more time for addressing changes, and the communication chain is longer. However, large organisations have more resources and budget to manage HCDs and likely have more diverse team members who may appreciate some HCDs than others. For example, survey respondents reported,
\textit{``Larger organisations would have different cultures involved and may make escalating to a relevant person harder, but they would have a wider base of experience'' - [SR-18].} \\
\textit{
``If the organisation is bigger there is more people who might be able to fix the problems faster'' - [SR-21].} While another survey respondent suggested, 
\textit{``This question again does not have one fixed answer. On one side bigger organisation with a huge number of diverse users can help us quickly identify different kinds of issues and can lead to fixing human-centric issues faster. Whereas a big organisation may also have a bigger chain of commands, getting approval on a problem or a ticket might take longer than usual due to a bigger hierarchy'' - [SR-3].}

Smaller organisations may lack the resources to handle or test some HCDs. They may also want to release a product faster to the market to gain a competitive advantage and, in turn, may lose focus on HCDs or de-prioritise some of them. However, the HCDs can be reported faster due to less chain of command and can sometimes be resolved more quickly. Some survey respondents claimed: \\
\textit{
``I do not think human-centric defects are considered by smaller organisations at all. Its not that the smaller organisations do not want to solve them; they just do not have the resources to test the products extensively.'' - [SR-13].\\
``The smaller the team, the quicker the bugs and dysfunctioning reach the developers'' - [SR-20].\\
``It depends on the response time of the organisation. The smaller the organisation, the faster bug may be reported to the developers and they might fix the issue faster. But sometimes developers may be working on some other project which is why they do not have time to fix the bug. This is where bigger organisations are better. Since they have more resources, it is easier to fix these bugs, but the downside is, it is difficult to report and get the bug to a stage where someone actually notices it and works on it'' - [SR-2].}

Sixteen percent of responses suggested that organisation size does not matter. They claimed if the teams are diverse, inclusive, and experienced enough, HCDs can be handled better. Additionally, if the developer follows good practices while developing software, it could be critical in handling HCDs. For instance, a survey respondent insinuated, \\
\textit{``Size of the organisation does not matter but by looking at the innovation and inclusive team in a company we can determine which organisations lead this case'' - [SR-10].} 
 Another survey respondent claimed, \textit{
``Based on my experience, I do not think it is the size of the organisation which causes these defects. But it may be due to the practices followed by the developers when developing the software. When the developers have different experience levels and work ethics, there is a chance different development practices can lead to human-centric defects'' - [SR-5].}

\subsubsection{Team climate impact on identifying and fixing HCDs}

Employees' collective impressions of organisational policies and processes are referred to as team climate \citep{anderson1998measuring}.
Based on the responses received in the survey, most respondents claimed team climate is essential in identifying and fixing of HCDs. For example, SR-10 mentioned, \textit{``Team climate plays a major role to allocate time and resources to human-centric defects and push for more fixing of defects''.} Team philosophy, diversity, and culture create an impact on the overall success of the product. If the team culture promotes a human-centric process, then HCDs can be fixed faster; otherwise, developers will always ignore blind spots and edge cases. For example, some survey respondents proposed: \\
\textit{
``The different cultures of the testers, developers and end-users has an impact on fixing and reporting of the human-centric issues.'' - [SR-18].\\
``Depending on the defect - a cultural blind-spot would affect the identification of the issue.'' - [SR-1].\\
``If the team culture is to prioritise human-centric issues, then these issues will get fixed faster'' - [SR-2].}

Team climate impacts the motivation, creativity, and productivity of a team delivering a product. The more diverse a group, the better the quality of the product. Additionally, it reduces communication delays and fastens the process of resolving issues. For example, SR-30 claimed that
\textit{
``A better team climate will generate more ideas, solutions and positive feedback. This will help fixing defects, but also identifying them more easily''.}

A shared perception can help focus the efforts, the issues to work on, and the issues to solve for better product quality. An organisation's cultural and sensitivity training could be a big part of the team's thinking while building software. For example, a survey respondent suggested, \\\textit{
``Mainly the attitude of the team, whether they are motivated or dissatisfied, is reflected in the possibility of finding defects'' - [SR-47].}

\begin{figure}[h]
  \includegraphics[scale=.4]{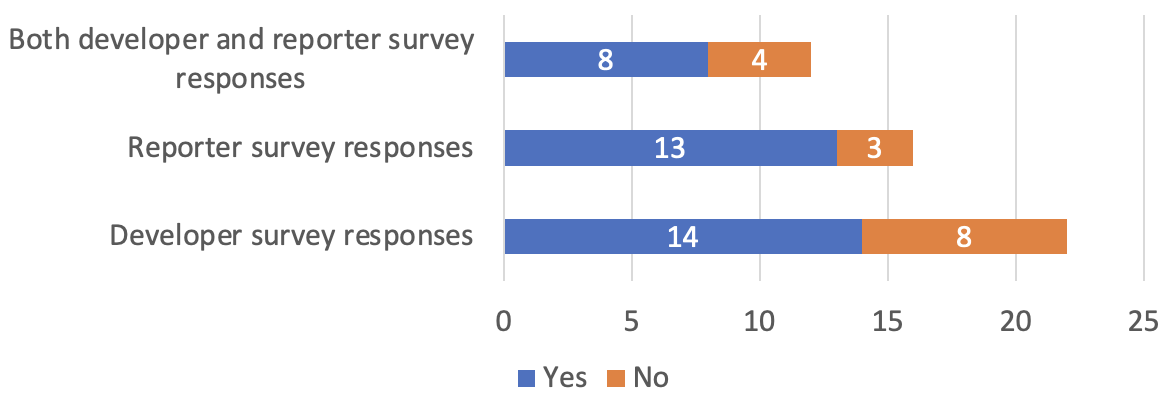}
\caption{Bar chart of geographic location impact in finding and fixing HCDs (in Surveys).}
\label{fig:barchartgeoloc}       
\end{figure}

\subsubsection{Geographic location's role in finding and fixing HCDs}


Based on the answers we received from the survey respondents as shown in Figure~\ref{fig:barchartgeoloc}, 70\% of respondents asserted that geographic location may add some delays in fixing and reporting of HCDs (or, defects in general). Distributed teams can be difficult to handle when the goal is to build a single standard software. There will be more discrepancies if the developers are located in different geographic locations because fixing a simple issue can take more hours due to varied working hours. For example, a survey respondent stated, \textit{``A defect identified in the USA time zone, has to wait for the India team to login and witness the issue. This can lead to wastage of hours to fix an issue'' - [SR-11].} In addition, communication across time zones restricts the progress and reporting of the problems. For instance, some survey respondents indicated: \\
\textit{``Access to resources and time play a huge role in finding defects. These factors can delay finding defects at different locations.'' - [SR-26]\\
``Offshore and timezone differences may lead to delay in resolving issues if there is no proper task management methodology or lack of communication.'' - [SR-3]\\}


Due to cultural and environmental differences, issues arising from different geographic areas can have translation problems. Native slangs could affect the bug terminology, making the HCDs longer to interpret. For example, a survey respondent claimed, \textit{``Translation issues are the biggest ``human centric'' defect as it can lead to delay in fixing issues'' - [SR-13].} The quality of reports can suffer due to spelling and grammar and even lead to the escalation of issues. Due to cultural differences finding a common solution for an HCD could also be a concern. For instance, a survey respondent stated, \textit{``Different locations have different ways of living, the culture might not be the same. So having a homogenous solution to a problem might serve one community while it might be detrimental to another. Those nuances in finding solutions definitely can delay a fix'' - [SR-37].}


\begin{tcolorbox}[arc=0mm,width=\columnwidth, top=1mm,left=1mm,  right=1mm, bottom=1mm,boxrule=1pt]
\small
    \textbf{RQ3 Key findings.} 

\begin{itemize}[leftmargin=*]
       \item Participants claimed that there needs to be more awareness about HCDs in organisations.
\item Diverse teams can expedite the process and help identify HCDs at earlier stages of the SDLC.
\item Organisation size could impact reporting and fixing of HCDs. Large organisations can be more bureaucratic and have time-consuming processes, but may have more resources to manage HCDs. Smaller organisations often lack such resources and may focus on quick-release cycles, but HCDs might be reported and resolved more swiftly, due to less chain of commands.
\item Team philosophy, diversity, and culture impact the product's overall success. If the team culture promotes a human-centric process, HCDs can be managed better. 
\item Majority of participants suggested that HCDs arising in one location and fixed in another could create significant delays. Due to cultural and environmental differences, issues arising from different geographic areas can have translation problems. 
\end{itemize}
\end{tcolorbox}

\subsection{RQ4 Results - Role of defect tracking tools in reporting and fixing HCDs}
\label{sec:rq4}

We identified that most of our participants have significant experience in using defect reporting tools. Table~\ref{table:defecttrackingtools} shows Jira and Trello\footnote{https://trello.com/} are actively used closed-source tools, and GitHub\footnote{https://github.com/} is the most actively used open-source tool by developers and reporters. 
The following sub-questions explore the existing defect-tracking tool handling of HCDs.

\begin{table*}[]
\caption{Popular defect tracking tools used by the participants}~\label{table:defecttrackingtools}
\begin{tabular}{@{}ll@{}}
\hline\noalign{\smallskip}
\textbf{Defect tracking tools used by the participants}      & \textbf{Overall} \\ \hline\noalign{\smallskip}
\textbf{Survey responses}                                    & \textbf{}        \\ \hline\noalign{\smallskip}
JIRA and GitHub                                              & 24.6\%           \\
Trello                                                       & 21\%             \\
Bugzilla                                                     & 11.4\%             \\
Zendesk                                                      & 9.6\%            \\
Service NOW                                                  & 7.9\%            \\
Redmine, Mantis, Remedy, Asana, Backlog, Zoho, BugHerd, etc. & 25.5\%           \\ \hline\noalign{\smallskip}
\textbf{Interview responses}                                 & \textbf{}        \\ \hline\noalign{\smallskip}
JIRA and GitHub                                              & 90\%             \\
In-house tool                                                & 10\%            \\ \hline\noalign{\smallskip}
\end{tabular}
\end{table*}

\subsubsection{RQ4a Results - Key challenges for defect tracking tools in reporting and resolving HCDs}
\label{sec:rq4a}

81.7\% of our practitioners claimed that defect tracking tools have a generic structure that does not provide many options to manage HCDs. One stated, \textit{``There is no option to handle human-centric issues in these tools. We usually create a regular Jira ticket without customisation'' - [IP-1].} 
Participants said that tools do not generally include (i) categories and tags for HCDs; (ii) no feature to associate end-user complaints with code commit history; (iii) the priority and severity of HCDs are inconsistent; (iv) lack of details in the reports such as steps to reproduce a problem and tool support; (v) time-consuming; (vi) task alignment and progression of the bugs lead to delays in fixing the defects. SR-10 claimed, 
\textit{``Defect reports provides limited tagging options on the defects, lack of categories and lineage for different types of defects''.} An interview participant stated,
\textit{``If you leave it to people to explain the bug detail, everyone would explain it differently'' - [IP-8].} 

18.3\% of participants consider these tools do not provide particular challenges because it depends on the reporter's skills in documenting the issue. 
An interview participant claimed,
\textit{``The skill of the reporter to accurately describe the issue is critical. The reporter that does not convey the problem is the major issue'' - [IP-7].} While some survey respondents suggested:\textit{``It comes to the reporter's skill in accurately capturing the defect. If an end user is unable/incapable of even submitting a complaint then this is a fundamental problem.'' - [SR-15]\\
``Most of the defect tracking tools are matured and have all the bells and whistles to capture the defect details'' - [SR-9].}

\subsubsection{RQ4b Results - Factors that make defect tracking tools user-friendly}
\label{sec:rq4b}

We asked our practitioners an open-ended question regarding possible recommendations for improving defect reporting tools to handle HCDs better. 
They said while creating a new defect in the tool, there should be an option to create more categories, tags, and sub-tasks relating to HCDs. In addition, there should be an alternative to providing feedback from a human perspective. IP-1 claimed that \textit{``There is a limitation in Jira to accommodate human-centric issues. The tool does not provide option to include feedback in various contexts such as contexts from a cultural or language perspective are not tracked.''.} The defect priority and severity should be adjusted based on the demographics impacted to reduce bias. Furthermore, the report should have the option to add a snapshot or video, a detailed and condensed description, and end-user surveys to make the defect precise for the developers. There should be defect report templates and targeted practices for more accessible learning and training to ensure human-centric requirements are not missed. Finally, they should include testing for edge conditions based on different perspectives to handle multiple scenarios while using a product. For instance, some survey respondents commented: \\
\textit{``By extending the workflow functionality and customisability to business teams,  they can track work, manage projects, and stay on top of everything.'' - [SR-26].\\
``By defining identified defects as tags, we can manage, and mitigate future human-centric issues.'' - [SR-10].}

\subsubsection{RQ4c Results - Key fields required in a defect reporting form to manage HCDs better}
\label{sec:rq4c}

\begin{figure}
  \includegraphics[scale=.5]{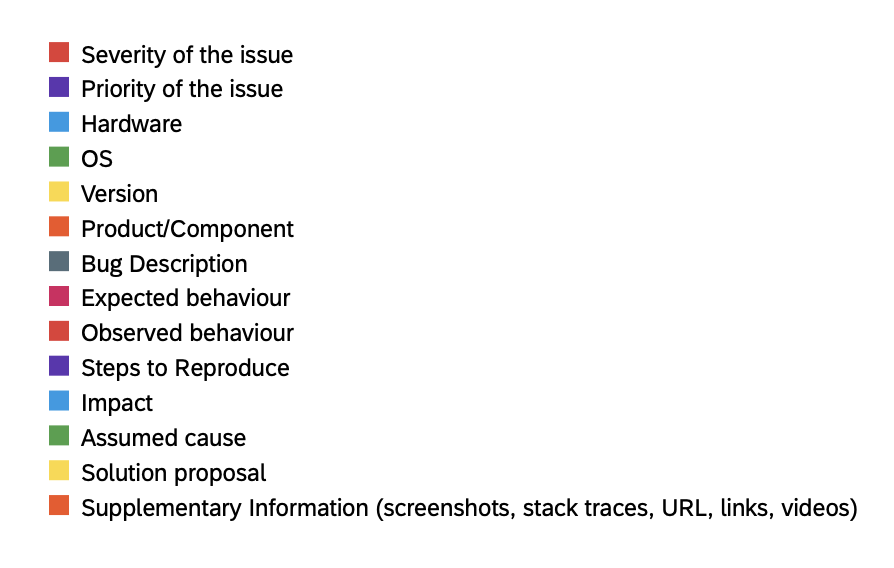}
\caption{Defect reporting fields}
\label{fig:defectrepfields}       
\end{figure}

We asked participants to common on a generalised defect reporting structure common in defect tracking tools such as Jira, Bugzilla\footnote{https://www.bugzilla.org/}, and GitHub. The commonly used fields in defect report forms are shown in Figure~\ref{fig:defectrepfields}.



\begin{figure*}
\centering
  \includegraphics[width=0.7\textwidth]{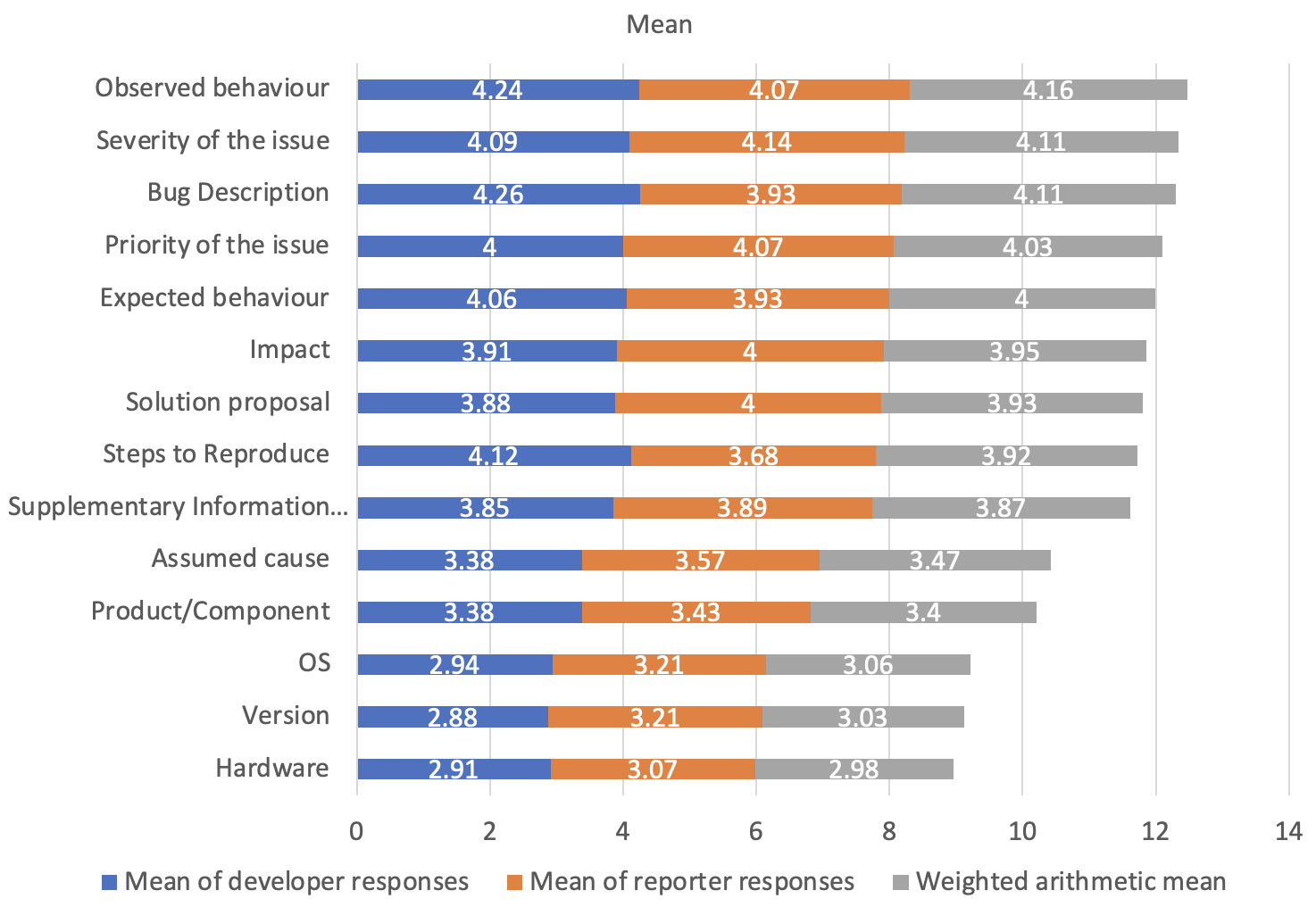}
\caption{Mean plot of defect reporting fields based on developer and reporter responses (in Surveys)}
\label{fig:meanplotdeffields}       
\end{figure*}

We asked respondents to rate these defect report fields based on their significance to understand and resolve HCDs. We used a Likert scale of 1 to 5, with `1' being the least important and `5' being the most significant. Figure~\ref{fig:meanplotdeffields} shows the mean plot of responses from developers and reporters based on the importance when creating an HCD. We used weighted arithmetic mean \citep{cochransampling} of developer and reporter responses as follows:
\begin{equation}
    x_{c} = \frac{m.x_{a} + n.x_{b}}{m+n}
\end{equation} 
where, 
\begin{itemize}
    \item x$_{\mathrm{a}}$ : the mean of the developer set,
    \item m : the number of developer responses (34),
    \item x$_{\mathrm{b}}$ : the mean of the reporter set,
    \item n : the number reporter responses (28),
    \item x$_{\mathrm{c}}$ : the weighted arithmetic mean.
\end{itemize}

The observed behaviour is highly rated, followed by the severity of the issue and bug description. This could be because HCDs are subjective and qualitative defects, and most of the tools provide an option for the observed behaviour, severity, and description, but a detailed report is fundamental. The least significant fields are Operating System (OS), version and hardware. 

\subsubsection{RQ4d Results - Influence of experience on HCD defect quality}
\label{sec:rq4d}


88\% of participants highlighted the role of experience in delivering high-quality HCD reports. 
Respondents suggested,
\textit{``To deliver high-quality product, we need to improve the Return on Investment (ROI) by reducing the cost of development. Overall, the communication, teamwork and connectivity should be better to detect issues earlier and understand defect trends. In short, the better, the service, the better is customer satisfaction which usually comes with experience'' - [SR-26].} \\ \textit{ 
``Experience in using a defect reporting tool will influence the quality of a defect report. The deeper you understand a tool and use all its features, you can generate better reports'' - [SR-3].}


Some respondents claimed that the formulation of a HCD defect report comes with experience and should improve over time. The tools are always secondary; the organisation must provide the necessary support and training to develop employees' knowledge base. SR-15 claimed, 
\textit{``A particular defect reporting tool is not much important - what is important is that there is organisational support to use and update them based on feedback''.}

\begin{tcolorbox}[ arc=0mm,width=\columnwidth,top=1mm,left=1mm,  right=1mm, bottom=1mm, boxrule=1pt]
\small
\textbf{RQ4 Key findings.} 
\begin{itemize}[leftmargin=*]
\item Jira and Trello are commonly used closed-source tools, and GitHub is the commonly used open-source tool.
\item Most practitioners highlighted that most current defect reporting tools provide a generic structure focusing on quantitative defects and don’t provide options to handle many HCDs, which are usually more qualitative. 
    \item Practitioners said that defect-tracking tools should provide snapshots or videos, a detailed and condensed description, end-user surveys, templates, and tags, to make the HCD reports more precise for developers.
    \item In a defect report for an HCD, fields such as observed behaviour, severity and bug description are fundamental. The least significant fields are OS, version, and hardware affected by the issue.
    \item A majority of practitioners claimed that experience in defect reporting tools could be influential in delivering high-quality defect reports. More experience can lead to better presentation and consistent reports. 
\end{itemize}
\end{tcolorbox}

\subsection{RQ5 Results - Improvements required for more effective reporting and resolution of HCDs}
\label{sec:rq5}


Practitioners said that \textbf{more automation} could be beneficial if we can learn from comments, version control histories, guidelines, and existing or previous issues to provide use cases to manage HCDs. The underlying guidelines for successful automation would be to avoid duplicate defects, aggregate different types of HCDs, learn from prior HCD causes and solutions, and improve the HCDs reporting process. For example, some survey respondents suggested: \\
\textit{``Learning and applying experience from past defects means you can fix them faster using the skills you have acquired. The same reasoning can be applied to human-centric defects.'' - [SR-16] \\
``I believe there is nothing more valuable then learning from old mistakes, report them, document them and use them in the future for faster work'' - [SR-27].} Some interview participants added to this by claiming: \textit{
``Automation would be beneficial to a good extent for aggregating all these defects such as scraping web to find similar human-centric issues.'' - [IP-5]\\
``A recommender engine which could create a knowledge base and taxonomy that underpins human-centric defects would be very useful.'' - [IP-2].}

The provision of an HCD-oriented \textbf{template or checklist while reporting issues} would help. A checklist on what characteristics will be beneficial for a developer to understand, appreciate and fix the HCD can help a reporter effectively capture information, speed up the process, and reduce delays. For example, an interview participant recommended, \textit{``In the case of the human-centric defects, a template should be there to guide the reporter on how to make a complete report'' - [IP-7].}

Some practitioners claimed that \textbf{more awareness} regarding HCDs is needed. One reason could be lack of \textbf{guidelines or taxonomy to understand HCDs}, similar to issues we found in reporting usability defects in prior work~\citep{yusop2016reporting,yusop2020revised}. If there were such a comprehensive list to help report, evaluate, and fix HCDs, then the management of HCDs in the SE domain could be better. For instance, IP-5 mentioned, \textit{``A set of standards for human-centric defects could benefit all developers. A list of things related to defects which developers can refer to and see what is applicable to them rather than waiting for the customer to provide them feedback and repeat the cycle''.} The taxonomy could categorise HCDs and test conditions based on culture, language, gender, age groups, educational background, and physical or mental disabilities. For instance, IP-7 stated, \textit{``The standards should contain human factors such as from colourblind perspective, make sure that the UI is interpretable''.} A broader taxonomy or a standard such as Design for All practices, Development practices, and W3C Accessibility Standards\footnote{https://www.w3.org/WAI/standards-guidelines/}, could be useful.


\begin{tcolorbox}[arc=0mm,width=\columnwidth,                 top=1mm,left=1mm,  right=1mm, bottom=1mm,boxrule=1pt] 
\small
\textbf{RQ5 Key findings.} \\
 To improve HCDs management, some suggestions are:
   \begin{itemize}[leftmargin=*]
       \item increased HCDs detection/classification automation;
        \item a standard taxonomy of HCDs for practitioners;
       \item improved HCD information capturing support in defect tracking tools.
    \end{itemize}
\end{tcolorbox}


\section{Discussion and Recommendations}
\label{sec:discussion}

\subsection{Improve HCDs' awareness and understanding}
There needs to be more awareness regarding HCDs to create a comprehensive working model to report, manage and solve such issues. Due to their subjectivity~\citet{huynh2021improving}, as confirmed in this research by multiple practitioners, different perceptions make it difficult to capture and understand HCDs. Every individual will perceive applications differently, and the usage criteria vary resulting in defects. Hence, awareness and understanding of HCDs are important in the current SE domain. 


\sectopic{Recommendation\#1.} 
Organisations and teams can create diverse, inclusive, and experienced environments to manage and promote understanding regarding HCDs efficiently.  HCDs should be actively considered as a part of the current defect management process to emphasise and highlight the problems diverse reporters face (Sections~\ref{sec:rq2}, \ref{sec:rq3}). 
\sectopic{Recommendation\#2.} 
Culture and sensitive training should be provided to all practitioners to understand different human aspects and their impact on defect reporting and fixing. \citet{yusop2016influences} stated that usability-related training to respondents leads to a better understanding of usability defects. Practitioners should be provided with templates and targeted practices for more accessible learning and training to ensure human-centric requirements are not missed (Sections~\ref{sec:rq2}, \ref{sec:rq3}). 


\sectopic{Recommendation\#3.}   Customer-centric processes should be created to help impacted end-users with a feedback loop to improve the existing processes. Co-creational spaces to enable software teams to capture critical human-centric requirements related to their software have been suggested by \citet{grundy2020humanise}. Organisations and teams can create a change management process based on end-users feedback, which can help highlight HCDs efficiently (Section~\ref{sec:rq3}).

\sectopic{Recommendation\#4.} Diverse user needs should be considered with solutions formulated around their needs. Identification practices such as trial and error, categorisation of issues, user's expected behaviour, accessibility issues, a recreation of user's approach and testing for edge conditions should be undertaken by teams. \citet{yusop2016reporting} provided key recommendations around usability defect attributes to capture attributes at a fine-grained level to support defect correction, which could be done for other HCDs. Teams need to cultivate an approach for focusing on HCDs in their development and testing practices (Section~\ref{sec:rq1}).

\sectopic{Research Directions.} Organisations and teams work on tight deadlines, and the resource constraint and pressure on practitioners to solve the issues without understanding the plights of their target audience results in more gaps between practitioners and end-users. This can lead to missing user requirements, providing a better experience to the users, bias towards your target audience, and overall a less engaging product. Researchers can focus on automation for generating requirements and system usage contexts from different perspectives. 
Furthermore, to be able to capture human-centric needs, there needs to be more research on formulating a taxonomy of HCDs. \citet{yusop2016reporting,yusop2020revised} suggests that usability defects suffer from an inconsistent taxonomy and recommend a taxonomy that can capture meanings and names of usability defect attributes to improve reporting, triaging, understanding, and fixing of these defects. To improve the management of HCDs in the software industry, a detailed taxonomy should be created. The taxonomy could categorise HCDs, provide better prioritisation of HCDs, improve defect report structure to accommodate HCDs, and test conditions based on culture, language, gender, age groups, educational background, and physical or mental disabilities (Section~\ref{sec:rq5}).


\subsection{Improve HCDs reporting and fixing}

\sectopic{Recommendation\#1.} HCD report structure needs to be improved by accommodating different categories, tags and information for different types of HCDs. The defect priority and severity should be adjusted based on the demographics impacted to reduce bias. \citet{yusop2016reporting} recommends a better severity, prioritisation, and custom reporting form to accommodate different types of usability defects. With regard to HCDs, different categories of HCDs need to be captured, reported and diagnosed to understand HCDs appropriately (Section~\ref{sec:rq4a}).

\sectopic{Recommendation\#2.} Customer information and feedback should be integrated into the HCDs tracking process. \citet{strate2013literature} suggests that defect report quality suffers from a disconnect between user reports and what the developers need in that report. It is also true for HCDs since HCDs could deviate depending on end-users and the report quality will also be dependent as not all users are proficient enough to provide quality reports. However, defect-tracking tools should provide basic features such as snapshots, videos, detailed descriptions, end-user surveys, templates, tags, and categories, to make the HCD reports more precise for the developer (Sections~\ref{sec:rq4b}, \ref{sec:rq4c}, \ref{sec:rq4d}).

\sectopic{Recommendation\#3.} Identifying and reporting HCDs is still highly manual. \citet{yusop2016influences,yusop2016reporting} suggests engaging automation in usability defect reporting tools could be more effective in capturing usability defects. Exploration of automated tools by introducing intuitive mechanisms to improve the process, detection, reporting, and fixing of HCDs could also be beneficial. For instance, if an HCD is reported, there could be automation to recognise the specific category of HCDs, triage multiple HCD reports for the same problem, and create a defect resolution process specific to that HCD. Similarly, analysing HCDs in relation to other defects can also lead to better identification and familiarity for the developers (Section \ref{sec:rq5}). 

\sectopic{Recommendation\#4.} Understanding and fixing HCDs can be better managed if a defect is assigned to domain experts with prior experience in HCDs. HCDs are subjective, and the understanding of defects can depend on the developer's empathy and understanding of human aspects leading to HCDs. This can be improved by organisations and teams employing diverse environments. \citet{strate2013literature} suggested automatic defect assignment based on the developer's background and commit history to reduce the time to fix the issue (Section~\ref{sec:rq1}). 


\sectopic{Research Directions.} Due to a lack of understanding about the aspects of different HCDs that are most critical to capture, and limited templates containing key HCDs characteristics to capture in current defect reporting tools, developers are not able to adequately capture and track HCDs. \citet{yusop2016reporting} recommends a wizard-based approach for reporting usability defects, especially for novice users. Specific HCD-oriented templates or checklists while reporting issues should also be investigated. However, with HCDs a checklist would need more comprehensive scoping from different human aspects to common end-user problems. Overall, a checklist will comprise what characteristics will be beneficial for a developer to understand, appreciate and fix the HCD. It can help a reporter in effectively capturing information as well as speeding up the process and reducing delays. Hence, a potential research avenue to capture and manage HCDs effectively (Section \ref{sec:rq5}).

\section{Study Limitations}
\label{sec:threats}


Whether participants comprehended the idea of HCDs the same way as we had intended poses the primary threat to internal validity. Results would be less trustworthy if practitioners did not comprehend the topics or used other meanings even though the practitioners have a working knowledge of defect management in the SE domain. To reduce participants' misunderstandings, we gave clear description and examples of each topic. We consider this threat to be minimal because we did not spot any misunderstandings or contradictions in the open-ended questions. Further, providing payment or gift cards to the participants in the second round of the survey and for interview could be another threat. We used two distinct crowdsourcing platforms, AMT and Prolific after considerations by similar studies \citep{jiarpakdee2021}. However, we authorised the payments for the respondents after reviewing their responses to determine if they answered each question and match our target participant group. 


Potential misinterpretation of questions is the main threat to construct validity. We developed our survey and interview questions with great consideration. We conducted pilot studies with software industry professionals and research professionals with previous software development and testing experience to provide clarity to the questions. Additionally, we provided practitioners with sufficient examples without creating bias so the responses provided by the practitioners could be based on their own judgement.



The limited survey size impacts generalisability. We received 104 responses, out of which 50 responses were selected. 
To mitigate the sample size of the survey, we conducted ten in-depth parallel interviews to understand the concepts of HCDs in SE. A community as diverse as software practitioners' cannot be quantified in terms of size. Therefore, it is probable that the sample used in our analysis does not accurately reflect the population as a whole. To mitigate this threat, we recruited participants from different countries and domains. 

Different organisations have different interpretations of the job roles of an individual. To reduce this threat, we interviewed developers and reporters with varied skills and experience. We gauged the skills of our respondents by rating their involvement in SE activities such as development, testing, database, usability, administration, and security. 


\section{Conclusion and Future Work}
\label{sec:conclusion}

HCDs occur in response to user perceptions or form inadvertently during the software development, therefore making it challenging to identify the problem and correct it before the products hit the market. Currently, development teams work on products based on a set of specified client requirements such as `what a system should do', i.e. functional, and `how it should do it', i.e. non-functional \citep{pandey2012importance,siddiqi1996requirements}. Unfortunately, the requirements often ignore the characteristics related to human-centric aspects such as personality, gender, emotions, language, engagement, ethnicity and culture, age, socioeconomic background, physical and mental challenges \citep{grundy2020human,hidellaarachchi2021effects}.  

This research explores the concept of human-centric aspects in defect reporting and resolution in the SE domain. To achieve the objective, we surveyed and interviewed software practitioners. We received fifty responses for a survey and conducted ten interviews to understand the perception of HCDs in SE. Based on the responses, we identified multiple areas of focus such as HCDs, identification and assignment of defects, experience role in defects, impact of human aspects on HCDs, organisation and teams role in HCDs, and management of HCDs in defect tracking tools. We identified that (i) HCDs are unique and are considered different from normal technical defects; (ii) the fixing and reporting of HCDs improve with experience; (iii) there is a serious lack of awareness of HCDs in organisations and teams that needs to be addressed; (iv) a diverse team of developers is more likely to address HCDs better; (v) the current defect reporting tools lack the capabilities to handle HCDs, which can be improved by automation support, and HCDs taxonomy and checklists for handling HCDs.

The field of defect tracking is large and handling human-centric aspects is still a major challenge. There are multiple avenues to explore such as better  discovery and triaging of HCDs, classification and categorisation of HCDs, and remediation and verification of HCDs. Based on the responses we gathered, our next direction is to explore existing detailed reporting processes and tools of HCDs. We want to study the challenges faced by diverse users while reporting their issues. Once we identify their challenges, we want to work on improving the reporting process and provide suggestions to enhance existing defect-tracking tools for better management and resolution of HCDs. This information will require close interaction with diverse users and ultimately can lead to best practices in reporting HCDs in the SE domain. We further plan to corroborate our findings from this study by analysing defect reporting and fixing in open-source repositories.


{\sectopic{Acknowledgment.} Chauhan and Grundy are supported by ARC Laureate Fellowship FL190100035.}\\

\sectopic{Declarations} 
{\sectopic{Conflict of Interest.} The authors declare that they have no conflict of interest.}
{\sectopic{Data Availability Statement.} The data is provided online}.

\printcredits

\bibliographystyle{cas-model2-names}

\bibliography{cas-refs}

\appendix
\section{Appendix}
The survey and interview questionnaires can be found on the Zenodo platform at https://doi.org/10.5281/zenodo.7978481.

\end{document}